\documentclass[usenatbib]{mnras}

\usepackage[T1]{fontenc}
\usepackage{ae,aecompl}

\usepackage{graphicx}	% Including figure files
\usepackage{amsmath}	% Advanced maths commands
\usepackage{amssymb}	% Extra maths symbols
\usepackage{nccmath}
\usepackage{gensymb}
\usepackage{caption}
\usepackage{xcolor}
\usepackage{soul}
\usepackage{enumitem}
\usepackage[normalem]{ulem}
\usepackage{tikz}
\usepackage{tikz-qtree}
\usetikzlibrary{positioning,shapes,snakes,calc}

\usetikzlibrary{positioning,shapes,snakes,calc}
\DeclareMathOperator{\Tr}{Tr}

\newcommand{\mh}{\mathrm{[M/H]}}
\newcommand{\ms}{\mathcal{M}}
\newcommand{\Prob}{\mathbb{P}}
\newcommand{\Se}{\mathrm{S}}

\newcommand{\bth}{\boldsymbol{\theta}}

\newcommand{\bep}{\boldsymbol{\epsilon}}

\title[Selection functions for spectroscopic MW surveys]{\texttt{seestar:} Selection functions for spectroscopic surveys of the Milky Way}

\author[Everall \& Das]{
Andrew Everall$^{1,2}$\thanks{Contact e-mail: \href{mailto:asfe2@cam.ac.uk}{asfe2@cam.ac.uk}},
Payel Das$^1$
\\
$^1$Rudolf Peierls Centre for Theoretical Physics, 1 Keble Road, Oxford, OX1 3NP, UK\\
$^2$Institute of Astronomy, University of Cambridge, Madingley Road, Cambridge, CB3 0HA, UK\\
}

\date{Accepted XXX. Received YYY; in original form ZZZ}

\pubyear{2018}

\begin{document}
\label{firstpage}
\pagerange{\pageref{firstpage}--\pageref{lastpage}}
\maketitle

% Abstract of the paper
\begin{abstract}
Selection functions are vital for understanding the observational biases of spectroscopic surveys. With the wide variety of multi-object spectrographs currently in operation and becoming available soon, we require easily generalisable methods for determining the selection functions of these surveys. Previous work, however, has largely been focused on generating individual, tailored selection functions for every data release of each survey. Moreover, no methods for combining these selection functions to be used for joint catalogues have been developed.

We have developed a Poisson likelihood estimation method for calculating selection functions in a Bayesian framework, which can be generalised to any multi-object spectrograph. We include a robust treatment of overlapping fields within a survey as well as selection functions for combined samples with overlapping footprints. We also provide a method for transforming the selection function that depends on the sky positions, colour, and apparent magnitude of a star to one that depends on the galactic location, metallicity, mass, and age of a star. This `intrinsic' selection function is invaluable for chemodynamical models of the Milky Way. We demonstrate that our method is successful at recreating synthetic spectroscopic samples selected from a mock galaxy catalogue.
\end{abstract}

% Select between one and six entries from the list of approved keywords.
% Don't make up new ones.

%%%%%%%%%%%%%%%%%%%%%%%%%%%%%%%%%%%%%%%%%%%%%%%%%%

%%%%%%%%%%%%%%%%% BODY OF PAPER %%%%%%%%%%%%%%%%%%

%%%%%%%%%%%%%
\section{Introduction}

Within our Milky Way, we are able to resolve stars out to large distances, but with that privilege comes a bias. The most comprehensive spectroscopic surveys can typically measure quantities for only a fraction of a percent of the stars in the Milky Way. The survey or `observed' selection function is the probability of a star being included in the survey given its sky positions, colour and apparent magnitude. To make inferences regarding the intrinsic chemodynamical structure of the Milky Way, knowledge is required of the intrinsic selection function, i.e. the probability of a star being included in the survey given its intrinsic coordinates: Galactic location, metallicity, mass, and age.

The second data release \citep{GaiaDR2} of the ESA'S Gaia mission has pushed us far further than ever before, measuring photometry, positions, and proper motions for over 1.3bn stars. In parallel, several ground-based spectroscopic surveys are measuring spectra for millions of these stars. Despite an often relatively simple nominal selection in colour and apparent magnitude, taking cross-matches with other surveys or selecting stellar type sub-samples result in a selection function that is no longer simple. Furthermore, to convert the observed colour-magnitude selection function into an intrinsic one that depends on distance, metallicity, mass, and age, one needs to engage with stellar isochrones. In order to understand the bias generated by selecting subsamples, we must calculate the selection functions for these surveys.

% Past studies
Many studies have developed survey selection functions, in which the completeness along a line of sight is given by the ratio of the number of stars in the spectroscopic survey to the number of stars in a photometric survey in the same region of colour and apparent magnitude. \cite{D16a} and \cite{D16b} use this method to construct selection functions for halo blue horizontal branch stars and K giants in Sloan Extension for Galactic Understanding and Exploration-2 \citep[SEGUE-II, ][]{SEGUE-BHB11}. Similar methods are used in determining the selection function for the Radial Velocity Experiment \citep[RAVE, ][]{K13} by \citet{W17} and for the Large Sky Area Multi-Object Fibre Spectroscopic Telescope \citep[LAMOST, ][]{LAMOST2012} Spectroscopic Survey of the Galactic Anticentre \citep[LSS-GAC, ][]{LSSGAC2-17} by \citet{Chen18}. \citet{Stonkute2016} perform a similar analysis of the Gaia-ESO survey \citep{GaiaESO12}, taking into account the survey's observing strategy.

\citet{Vickers18} generate a combined selection function for LAMOST \citep{LAMOST12}, RAVE \citep{K17} and the Tycho-Gaia Astrometric Solution (TGAS, \citealp{TGAS15}) by binning the fields in colour, apparent magnitude, and distance using a synthetic galaxy catalogue as the assumed complete sample. \citet{Bovy12} and \citet{Bovy14} consider the dependence of the selection function on apparent magnitude for G-type dwarfs in the Sloan Extension for Galactic Understanding and Exploration \citep[SEGUE, ][]{SEGUE09} survey and the Apache Point Observatory Galaxy Evolution Experiment \citep[APOGEE, ][]{Majewski17}. Their selection function along any line of sight is assumed to be uniform in apparent magnitude with limits defined either by the faintest star observed in a field or by the survey's nominal magnitude limit. \citet{Bovy14} also present a selection function in distance that takes into account the dust extinction along the line of sight.

Finally, \citet{Nandakumar17} and \citet{Mints19} determine selection biases for a large number of spectroscopic surveys by binning the sample in colour-magnitude space, either using a regular grid or a specialised median binning algorithm, and comparing with the 2-Micron All Sky Survey (2MASS, \citealp{Sk6}).

Amongst all previous works, fundamental aspects of the methodology remain the same. The dependence of the selection function on colour and magnitude is treated as a ratio of number counts of stars between the spectroscopic and a given photometric survey or a synthetically generated population, which is assumed to be complete. Uncertainties in measurements in colour and apparent magnitude are not used in the calculation. The effects of overlapping coordinate fields on the selection function are not considered. Selection functions are largely constructed as a function of colour and magnitude and not converted to intrinsic coordinates, with the exception of \cite{D16a}, \cite{D16b}, and \cite{San15}. With the exception of \citet{Vickers18}, no methods are presented for combining selection functions.

%This paper
In this paper, we build on previous work to create \texttt{seestar}\footnote{https://github.com/aeverall/seestar}, a {\sc Python} code that can be applied to any multi-fibre spectroscopic survey, independent of its footprint on the sky, the number of stars observed, and the selection criteria of the survey. We construct an algorithm to treat the limitations of Poisson noise when bins have small numbers of stars by using a Poisson point process, whose parameters we determine using maximum likelihood estimation. We propose how the uncertainties in colour and magnitude measurements may be incorporated into this. We include a method for calculating the union of overlapping field probabilities, which also enables selection functions of independent surveys to be combined. We use isochrones to convert the selection function depending on colour and apparent magnitude to one that depends on distance, metallicity, mass, and age. This is an essential component of chemodynamical models of the Milky Way \citep{Sch09,San15,D16a}.

In Section~\ref{sec:method}, we demonstrate how to calculate the selection function in observable coordinates (colour and apparent magnitude) and intrinsic coordinates (distance, metallicity, mass, and age), and how to determine the dependence on sky position. The results of tests on a mock catalogue are presented in Section~\ref{sec:mock}. Finally, we discuss our results and potential future developments and applications of this method in Section~\ref{sec:discussion}. Section~\ref{sec:method} is rather technical in nature, and we therefore advise readers only interested in the performance of the selection function package to skip to the results.

Appendix~\ref{app:modelapp} provides a table of parameter descriptions for reference along with a graphical Bayesian network to help with following our notation and understanding the method.

%%%%%%%%%%%%%%%%%%%%%%%%%
 \section{Method}\label{sec:method}

The selection function $\Prob(\Se \mid  \mathbf{x})$ of a stellar survey is the probability of a star being in the survey, $\Se$, given the star's coordinates $\mathbf{x}$. The coordinates may be observed (sky positions, apparent magnitude, and colour) or intrinsic quantities (galactic location, metallicity, mass, and age).
We assume that for the spectroscopic survey, there is a photometrically selected catalogue of stars which is complete in the region of colour and apparent magnitude observable by the spectrograph. Throughout this paper, we will use the superscript `spec' to refer to the spectroscopic sample and `phot' to refer to the photometric sample.
By Bayes' theorem
\begin{equation}
\label{eq:sfdef}
\Prob(\Se \mid   \mathbf{x}) = \frac{ \Prob(\mathbf{x} \mid  \Se)   \Prob(\Se) } { \Prob( \mathbf{x} )}.
 \end{equation}

$\Prob(\mathbf{x}) \mathrm{d}^n x \equiv f(\mathbf{x}) \mathrm{d}^n x$ is the probability that a star chosen at random has coordinates within the volume element, $\mathrm{d}^nx$, where $f(\mathbf{x})$ is the distribution function of stars in the Milky Way. $\Prob(\mathbf{x}   \mid  \Se)  \mathrm{d}^nx \equiv f^\mathrm{spec}(\mathbf{x}  \mid  \Se)  \mathrm{d}^nx$ is the probability that a star picked at random from the survey has coordinates in $\mathrm{d}^nx$, where $f^\mathrm{spec}(\mathbf{x})$ is the distribution function of observed stars. Finally, $\Prob(\Se) = \mathbb{E}\left[N^\mathrm{spec}\right] / \mathbb{E}\left[N\right]$ is the probability of a star entering the survey and is given by the ratio of the expected number of stars in the survey to the expected total number of stars in the Milky Way. Under the assumption that the photometric sample is complete within the given range of observable coordinates, we denote $f^\mathrm{phot}(\mathbf{x}) = f(\mathbf{x})$ and $N^\mathrm{phot}=N$ and substitute into Equation~\eqref{eq:sfdef}.
\begin{equation}
\label{eq:sfsmooth}
\Prob(\Se \mid   \mathbf{x}) = \frac{ f^{\mathrm{spec}}(\mathbf{x}) \mathbb{E}\left[N^{\mathrm{spec}}\right]} { f^\mathrm{phot}(\mathbf{x}) \mathbb{E}\left[N^\mathrm{phot}\right]} =
\frac{ \mathbb{E}\left[n^\mathrm{spec}(\mathbf{x})\right]}{ \mathbb{E}\left[n^\mathrm{phot}(\mathbf{x})\right]},
\end{equation}
where $\mathbb{E}\left[n^\mathrm{spec}(\mathbf{x})\right]$ and $\mathbb{E}\left[n^\mathrm{phot} (\mathbf{x})\right]$ are the expected number densities of stars in the spectroscopic and photometric surveys respectively, at coordinates $\mathbf{x}$.

The observation of a star is dependent on the star's observable properties, $\mathbf{x} = (l, b, c, m)$ where $(l,b)$ are the star's coordinates on the sky and $(c, m)$ the colour and apparent magnitude of the star. For ease of notation we will group these into positional coordinates, $\bth = (l, b)$ and photometric properties, $\mathbf{v}=(c, m)$.

The positional coordinates indicate which region of the sky or `patch' the star belongs to. The best method to characterize the dependence of the selection function on these patches depends on the survey design. This is described in Section~\ref{sec:skypos}. On each patch, the selection function is calculated as a function of $\mathbf{v}$ as described in Section~\ref{sec:obssf}. In Section~\ref{sec:sfint}, the observed coordinates are transformed into intrinsic coordinates, $\mathbf{v}(s, \mh, \mathcal{M}_{\mathrm{ini}}, \tau)$, where $s, \mh, \mathcal{M}_{\mathrm{ini}}, \tau$ are distance, metallicity, initial mass, and age respectively. These coordinates underpin the description of the chemodynamical distribution of stars in the Milky Way.

%-------------------
\subsection{Dependence of selection function on sky positions}
\label{sec:skypos}

To determine the dependence of the selection function on positional coordinates, we bin the sky into independent regions called patches across which the selection function does not vary directly as a function of sky position. Therefore we can denote $\Theta_i$ as the event that the star's coordinates are located on patch $i$. Using this notation, we can define the probability of a star being located on a field given the star's sky positions as
\begin{equation}
\label{eq:fieldcases}
\Prob(\Theta_i  \mid  \bth) =
\begin{cases}
1 &  \text{if $\bth$ is located on patch $i$},\\
0 &   \mathrm{otherwise}.
\end{cases}
\end{equation}
%

%==========================
\subsubsection{Defining a patch}
\label{sec:defpatch}

We assume that the selection function given $\mathbf{v}$ is independent of $\bth$ across any patch. This motivates us to divide the sky into smaller patches particularly in locations where we expect rapid changes in the selection function with respect to sky positions. However, using smaller patches reduces the number of stars per patch, which amplifies the effects of Poisson noise. A sufficient coarseness is required such that the effects of Poisson noise are limited but fine enough such that the dependence of the selection function on sky positions is small.

Before continuing, we should briefly explain what we mean by Poisson noise and why it is so central to this work. Detections and recordings of stars in stellar catalogues can be considered as independent events which are randomly sampled from a distribution over observed coordinates. Thus the detection of stars by stellar surveys is well modelled by a Poisson point process. The signal-to-noise ratio for a Poisson point process scales with the square root of the mean signal strength, $\frac{\mathrm{S}}{\mathrm{N}} \sim \sqrt{\lambda}$ where $\lambda$ is the expected number of events per interval of observed coordinate space. Therefore the larger the number of events per interval, the stronger the signal-to-noise ratio. This is why reducing the number of stars per patch in order to increase the spatial resolution also leads to a reduction in the signal-to-noise ratio.

\begin{figure}
\captionsetup{justification=centering}
       \includegraphics[width = \columnwidth]{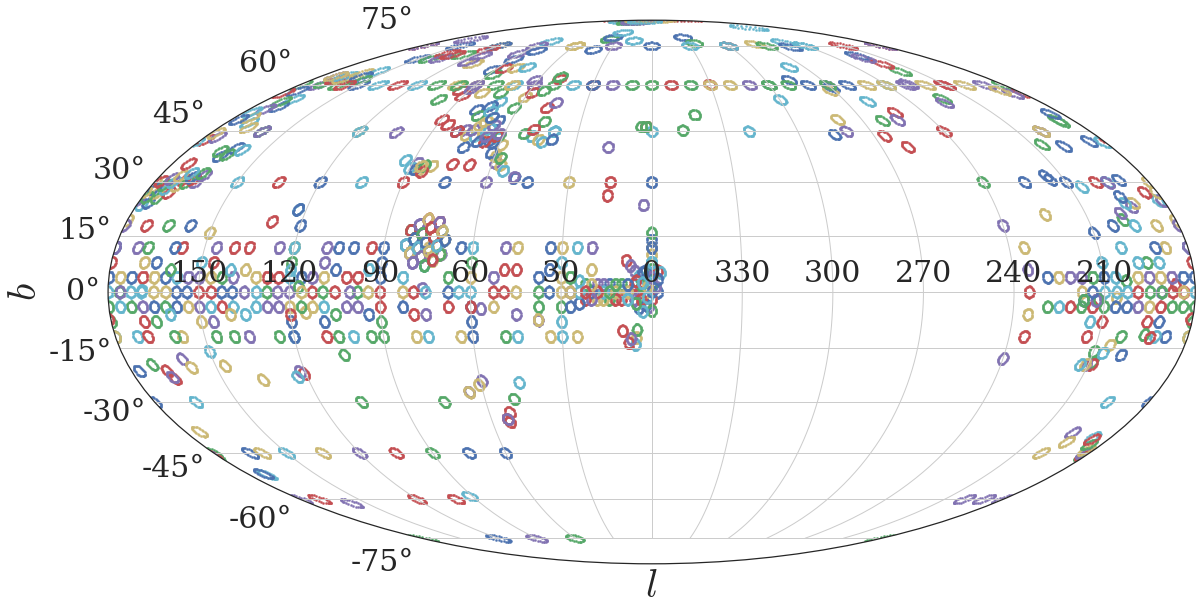}
      \includegraphics[width = \columnwidth]{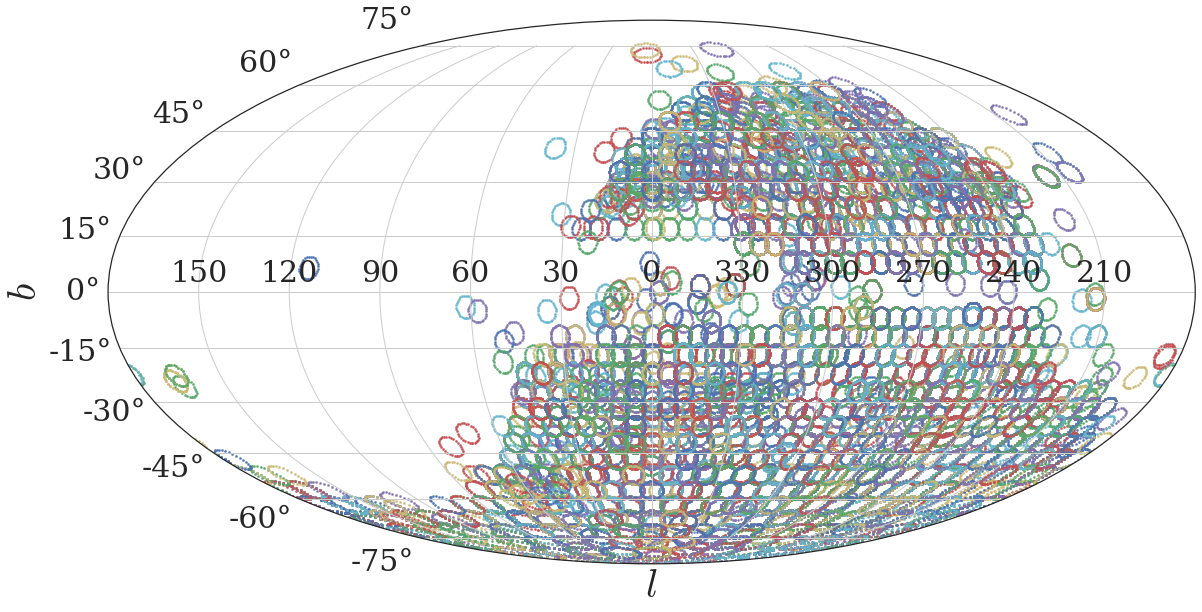}
\caption{Individual fields in APOGEE DR14 (top) and RAVE DR5 (bottom), shown here in galactic coordinates, have substantial overlap within each survey. Despite being northern and southern hemisphere surveys respectively, their footprints overlap particularly in the equatorial plane.}
\label{fig:fieldsfootprint}
\end{figure}

A multi-fibre spectroscopic survey is constructed by placing fibres on a plate so that each fibre is at the image of a star. The plate covers a solid angle on the sky. The solid angle of sky observed by each plate is here referred to as a {\it field}, and the coordinates of the centre of the field define the {\it field pointing}. The distributions of fields for APOGEE and RAVE in sky positions are shown in Figure~\ref{fig:fieldsfootprint} where each coloured circle represents the extent of a different field, and the field pointings are located at the centre of each circle. For multi-fibre spectroscopic surveys we define a patch as the region of sky covered by a field. From here on we will refer to patches as fields.

For a point on a single field, the selection function is the probability that the star will be observed by that field given the star's photometric properties, $\mathbf{v}$. If a coordinate does not lie in a field, the selection function is $0$.

The selection function for field $i$ is given by
\begin{equation}
\label{eq:fieldselection}
\Prob(\Se_i\mid  \bth, \mathbf{v}) = \Prob(\Se_i \mid  \Theta_i, \mathbf{v})\Prob(\Theta_i \mid  \bth)
\end{equation}
where $\Se_i$ is the event that a star is selected by field $i$. 

In many spectrograph surveys, as can be seen for APOGEE and RAVE in Figure~\ref{fig:fieldsfootprint}, fields may heavily overlap leading to a single coordinate being observed multiple times on different fields. The selection function is the probability that \textit{at least} one field selects the star. This is given by the union of the event of each field selecting the star.

\begin{equation}
\label{eq:fullunion}
\Prob(\Se \mid \bth, \mathbf{v} )  = \Prob\left(\bigcup_{i=1}^M\Se_i\,\bigg|\,\bth, \mathbf{v} \right)
\end{equation}

where $M$ is the total number of fields employed. This is calculated using all observations made on field $i$, even if the same star is observed on another field. The probability of the union of being on either of the two fields is the probability that one or the other occurs. We expand this in terms of the selection by individual fields

\begin{equation}
\begin{split}
\label{eq:fullunionexpanded}
\Prob \left (\bigcup_i^M \Se_i\,\bigg|\,\bth, \mathbf{v} \right) = \sum_{k=1}^M (-1)^{k+1} \left[   \sum_{1\leq i_1 < ... < i_k \leq M} \Prob(\Se_{i_1}, \Se_{i_2} ...\Se{i_k}) \right]
 \end{split}
\end{equation}
where 
\begin{equation}
\begin{split}
\label{eq:jointprobability}
\Prob(\Se_1, \Se_2, ...) = \prod_{i = 1, 2, ...}   \Prob(\Se_i \mid  \Theta_i, \mathbf{v}) \Prob(\Theta_i \mid  \bth).
\end{split}
\end{equation}

assuming the events $\Se_1, \Se_2...$ are independent. Appendix~\ref{app:unionexp} provides a detailed explanation of this expansion.

As mentioned earlier, many surveys contain multiple observations of the same positional coordinates. These can be observations taken on separate days to different magnitude depths. For Equation~\eqref{eq:jointprobability} to be appropriate, different observations of the same field should only be considered as separate fields if the observations are independent, i.e. if the probability of a star being selected by one observation is independent of whether the star is selected by the other observation. If the observations are dependent, as is the case if stars in one observation are chosen deliberately to be exclusive or inclusive of those observed in another observation, then the observations should be combined to form a single field in the selection function. In our method we combine all observations with the same field pointing as a single field.

In the following sections, we examine the dependence of the selection function on $\mathbf{v}$. These are calculated for a given field, $i$. Having calculated the selection function for each field as a function of $\mathbf{v}$, they are combined using the method above to achieve the full selection function for the entire survey.

%--------------
\subsection{Selection function in observable coordinates}
\label{sec:obssf}

The probability of a star being selected by field $i$ given $\mathbf{v}$, and given that the star lies on the field (i.e. $\Prob (\Theta_i\mid\bth) = 1$) is
\begin{equation}
\label{eq:isfdef}
\Prob(\Se_i\mid \mathbf{v},\Theta_i) = \frac{\Prob(\mathbf{v}\mid\Se_i, \Theta_i)   \Prob(\Se_i\mid\Theta_i)}
{\Prob(\mathbf{v} \mid\Theta_i) }.
\end{equation}
$\Prob(\mathbf{v} \mid\Theta_i) \mathrm{d}\mathbf{v} \equiv f (\mathbf{v} \mid \Theta_i) \mathrm{d}\mathbf{v}$ is the probability that a star chosen at random on field $i$ has coordinates within the $\mathrm{d}\mathbf{v}$ volume element, where $f(\mathbf{v}\mid\Theta_i)$ is the distribution of stars in the Milky Way inside the cone projecting onto field $i$.

$\Prob(\mathbf{v} \mid\Se_i, \Theta_i) \mathrm{d}\mathbf{v} \equiv f^\mathrm{spec}(\mathbf{v} \mid\Theta_i)  \mathrm{d}\mathbf{v}$ is the probability that a star observed on field $i$ of the spectroscopic survey has coordinates within the $\mathrm{d}\mathbf{v}$ volume element. Finally, $\Prob(\Se_i) = \mathbb{E}\left[N_i^{\mathrm{spec}}\right] / \mathbb{E}\left[N_i\right]$ is the probability of a star on field $i$ entering the survey, and is given by the ratio of the number of stars in the survey on field $i$ to the total number of stars in the Milky Way inside the cone projecting onto field $i$. Assuming that the photometric sample is complete within the given range of observable coordinates $f^\mathrm{phot}(\mathbf{v}\,\mid\,\Theta_i) = f(\mathbf{v}\,\mid\,\Theta_i)$ and $N_i^\mathrm{phot} = N_i$. Substituting into Equation~\eqref{eq:isfdef}
\begin{equation}
\label{eq:isf2}
\Prob(\Se_i\mid\Theta_i,\mathbf{v}) = \frac{f_i^{\mathrm{spec}} (\mathbf{v}\mid\Theta_i)\mathbb{E}\left[N_i^{\mathrm{spec}}\right]}
									{f_i^{\mathrm{phot}}(\mathbf{v}\mid\Theta_i) \mathbb{E}\left[N_i^{\mathrm{phot}}\right]}
= \frac{\mathbb{E}\left[n_i^{\mathrm{spec}}(\mathbf{v}\mid\Theta_i)\right]}{\mathbb{E}\left[n_i^{\mathrm{phot}}(\mathbf{v}\mid\Theta_i)\right]},
\end{equation}
where $n_i^{\mathrm{spec}}(\mathbf{v}\mid\Theta_i)=f_i^{\mathrm{spec}} (\mathbf{v}\mid\Theta_i)N_i^{\mathrm{spec}}$ is the number density of stars observed by the survey on field $i$ and $n_i^{\mathrm{phot}}(\mathbf{v}\mid\Theta_i)=f_i^{\mathrm{phot}} (\mathbf{v}\mid\Theta_i)N_i^{\mathrm{phot}}$ is the number density of stars in the Milky Way on the cone projecting onto field $i$.\\

%============
\subsubsection{Number density of photometric sample}
\label{sec:dfobs}

We start by calculating the expected number density of stars in the Milky Way on field $i$, $\mathbb{E}\left[n_i^{\mathrm{phot}}(\mathbf{v}\mid\Theta_i)\right]$. The choice of photometric survey is discussed in Section~\ref{sec:discussion}.\\

The stars in the photometric survey represent a Poisson realisation of the smooth underlying number density function, $n^\mathrm{phot}$. The aim is to use the observed stars to estimate the true smooth number density function. We assume this function can be parameterised as a bivariate Gaussian Mixture Model (GMM) for each field $i$. Each bivariate Gaussian component is given by
\begin{equation}
G(\mathbf{v} \mid \boldsymbol{\mu}, \boldsymbol{\Sigma}) = \frac{1}{\sqrt{\mid2\pi\boldsymbol{\Sigma}\mid}}
\exp\left(-\frac{ (\mathbf{v}-\boldsymbol{\mu})^T  \boldsymbol{\Sigma}^{-1}  (\mathbf{v}-\boldsymbol{\mu})}{2}\right),
\end{equation}
where $\boldsymbol{\mu} \in \mathbb{R}^2$ is the mean of the bivariate Gaussian in colour and magnitude and $\boldsymbol{\Sigma} \in \mathbb{R}^{2\times2}$ is the symmetric covariance matrix. Using this, we can write
\begin{equation}
n_i^{\mathrm{phot}}(\mathbf{v} \mid \bep_i,\Theta_i) = \sum_{k=1}^{K} w_{i k}   G(\mathbf{v} \mid \boldsymbol{\mu}_{i k}, \boldsymbol{\Sigma}_{i k}),
\end{equation}
where $\bep_i \in \mathbb{R}^{K\times 6}$ for the $K$ components of the GMM, each with six parameters defining the bivariate Gaussian component $w$, $\mu^0$, $\mu^1$, $\Sigma^{00}$, $\Sigma^{11}$ and $\Sigma^{01}$ ($\boldsymbol{\Sigma}$ is symmetric so $\Sigma^{10}=\Sigma^{01}$).

We then find the parameters $\bep_i$ by maximising the likelihood of the photometric catalogue stars. The Poisson likelihood is derived from Poisson count probabilities to give
\begin{equation}
 \begin{split}
   \label{eq:poisloglike}
  \ln(\mathcal{L}^\mathrm{phot}(X_i^\mathrm{phot} | \bep_i)) = -\int &\mathrm{d}\mathbf{v}\, n_i^\mathrm{phot}(\mathbf{v}\mid\bep_i, \Theta_i) \\
  &+ \sum_{j=1}^{N_i^{\mathrm{phot}}} \log \left( n_i^{\mathrm{phot}}(\mathbf{v}_j \mid \bep_i, \Theta_i) \right), \\
  \end{split}
  \end{equation}
where $\mathbf{v}_j$ are the colour and apparent magnitude of star $j$ in the photometric catalogue on field $i$ (see Appendix~\ref{app:poislike}).

We reparameterise the weights as normalised `loadings', $\pi_{i k} = w_{i k}/N$ where $N = \sum_{k=1}^K \left[w_{i k}\right]$. 
\begin{equation}
\begin{split}
\label{eq:normpoisloglike}
\ln (\mathcal{L}^\mathrm{phot}(X_i^\mathrm{phot} | \bep_i)) \propto - &N + N^\mathrm{phot}_i\ln(N) \\
&+\sum_{j=1}^{N^\mathrm{phot}_i} \ln\left(\hat{n}^\mathrm{phot}_i (\mathbf{v}_j \mid \bep_i, \Theta_i)\right),
 \end{split}
\end{equation}
where $\hat{n}^\mathrm{phot}_i = n^\mathrm{phot}_i/N$. The normalisation, $N$ is now a parameter of the model whilst the loadings provide $K-1$ free parameters under the constraint $\sum_{k=1}^K \pi_{i k} = 1$.

To obtain the posterior probabilities on each of our parameters we need priors. 
The prior on the normalisation $N$ is uniform, $N \sim U\left[0, \infty\right]$.
For the mean and covariance of the Gaussian components, we employ a Normal Inverse Wishart (NIW) prior.
%\frac{1}{\sqrt{|\Sigma|}} \exp{\left[ (\mu - m_0)^T \Sigma^{-1} (\mu - m_0)\right]}
\begin{equation}
\begin{split}
\mathrm{NIW}(&\boldsymbol{\mu}, \boldsymbol{\Sigma} \,\mid\, \mathbf{m}_0, \lambda, \boldsymbol{\Psi}, \nu) \\
\propto & \quad \mathcal{N}\left(\boldsymbol{\mu} \mid \mathbf{m}_0, \frac{1}{\lambda}\boldsymbol{\Sigma} \right)  
|\boldsymbol{\Sigma}|^{-\frac{(\nu + p + 1)}{2}} \exp{\left[-\frac{1}{2} \Tr{\left(\boldsymbol{\Psi} \boldsymbol{\Sigma}^{-1}\right)}\right]},
\end{split}
\end{equation}
where $p$ is the number of degrees of freedom of the system, in our case two (colour and apparent magnitude). The Normal and Inverse Wishart distributions are the conjugate priors for the mean and covariance of a multivariate Gaussian distribution respectively. As a result, the posterior on $\boldsymbol{\mu}$ and $\boldsymbol{\Sigma}$ will also be represented by a NIW distribution. The chosen hyperparameters ($\mathbf{m}_0, \lambda, \boldsymbol{\Psi}, \nu$) are dependent on the prior information about the data. Choosing small values of $\lambda$ and setting $\nu = p$ provides a least informative prior on the parameters. We discuss a good choice of hyperparameters in our mock tests in Section~\ref{sec:mock}.
 
The loadings are evaluated with a Dirichlet prior,
\begin{equation}
\mathrm{DIR}(\pi_1, \pi_2...\pi_K | \alpha_1, \alpha_2...\alpha_K) \propto \prod_{k=1}^{K} \pi_k^{\alpha_k - 1}
\end{equation}
which satisfies the constraint $\sum_{k=1}^K \left[\pi_{i k}\right] = 1$.
The Dirichlet distribution is the conjugate prior for membership probability of a categorical variable. Setting concentration hyperparameters $\alpha_i = \alpha = 1/K \,\forall\, i$ provides a least informative prior.

For a given  number of components, $K$ we determine the posteriors on parameters of the normalised GMM using the {\sc BayesianGaussianMixture} module from {\sc scikit-learn} \citep{sklearn} with parameters initialised by $k$-means clustering. This module iterates using a Variational Inference algorithm which alternates between assigning component membership probabilities and fitting the individual components.
We recover the best fit hyperparameters of the posterior Dirichlet and NIW distributions $\alpha_k', \mathbf{m}_{0k}', \lambda_k', \boldsymbol{\Psi}_k'$ and  $\nu_k'$ where $k$ is the Gaussian mixture component. These hyperparameters define the posterior distribution of parameters for the GMM.
The posterior on the normalisation, $N$ is given by the first two terms of the right hand side of Equation~\eqref{eq:normpoisloglike} where the uniform prior doesn't provide a contribution. This is proportional to a Poisson distribution with mean $N^\mathrm{phot}_i$.

To determine the number of components, $K$ we fit our GMM with $K=1$ up to a maximum of $K=20$ components and calculate the Bayesian information criterion \citep[BIC, ][]{Schwarz1978} for the best fit parameters (taken as the expected values of the posteriors)
\begin{equation}
\mathrm{BIC} =\ln(N_i^\mathrm{phot})d - 2\ln(\mathcal{L}^\mathrm{phot}_\mathrm{max})
\end{equation}
where $d=K\times 6$ (the number of parameters in our model) and $\mathcal{L}^\mathrm{phot}_\mathrm{max}$ is the likelihood corresponding to the best fit parameters. Our final model for the photometric sample colour-apparent magnitude function is the model which minimizes the BIC.

%
%============
\subsubsection{Selection function in observed coordinates}
\label{sec:sfobs}

\begin{table*}
\centering
\begin{tabular}{|| c | c | c | c  | c | c || }
\hline
Data & Component & Model & Parameters & Prior & Hyperparameters\\ 
 \hline
Photometric & Distribution Function & GMM & $N_i$ & $\mathrm{U}[0, \inf]$ &\\ 
& 	&	&	&	&\\
&   &   & $\pi_{ik}$ & Dirichlet & $\alpha = 1/K$ \\ 
& 	&	&	&	&\\
&   &   & $w_{i k} = N_i\,\pi_{i k}$  &   &\\
& 	&	&	&	&\\
&  	&	& $ \boldsymbol{\mu}_{ik}$, $\boldsymbol{\Sigma}_{ik}$ & NIW & $\mathbf{m}_0=\mathrm{med}[\mathbf{v}_{\mathrm{min}}, \mathbf{v}_{\mathrm{max}}]$, $\lambda=10^{-4}$,\\ 
& 	&	&	&	& $\boldsymbol{\Psi}=\left(\frac{\mathrm{diag}(\mathbf{v}_\mathrm{max}-\mathbf{v}_\mathrm{min})/2}{5}\right)^2$, $\nu=2$\\
 \hline
Spectroscopic & Distribution Function & GMM & $N_i$	& Poisson	& $\mathrm{mean} = N_i^{\mathrm{phot}}$\\
& 	&	&	&	&\\
& 	&	& $\pi_{ik}$ & Dirichlet & $\alpha_{ik}'$  \\
& 	&	&	&	&\\
&   &   & $w_{i k} = N_i\,\pi_{i k}$  &   &\\
& 	&	&	&	&\\
& 	&	& $ \boldsymbol{\mu}_{ik}$, $\boldsymbol{\Sigma}_{ik}$ & NIW & $\mathbf{m}_{0ik}'$, $\lambda_{ik}'$ \\
& 	&	&	&	& $\boldsymbol{\Psi}_{ik}'$, $\nu_{ik}'$\\
& 	&	&	&	&\\
& Selection Function & GMM & $\widetilde{\kappa_{ik}}$ & $\mathrm{U}[-\inf, \inf]$  &  \\
& 	&	&	&	&\\
&   &   & $\widetilde{w_{i k}} = \sqrt{|2\pi\widetilde{\boldsymbol{\Sigma}_{ik}}|}\,\mathrm{logit}^{-1}(\widetilde{\kappa_{ik}})$  &   &\\
& 	&	&	&	&\\
& 	&	& $ \widetilde{\boldsymbol{\mu}_{ik}}$, $\widetilde{\boldsymbol{\Sigma}_{ik}}$ & NIW & $\widetilde{\mathbf{m}_0}=\mathrm{med}[\mathbf{v}_{\mathrm{min}}, \mathbf{v}_{\mathrm{max}}]$, $\widetilde{\lambda}=10^{-4}$,\\ 
& 	&	&	&	& $\widetilde{\boldsymbol{\Psi}}=\left(\frac{\mathrm{diag}(\mathbf{v}_\mathrm{max}-\mathbf{v}_\mathrm{min})/2}{5}\right)^2$, $\widetilde{\nu}=2$\\
 \hline
\end{tabular}
\caption{Parameters and hyperparameters of the GMMs and their respective priors to be fit to the photometric and spectroscopic catalogues. $i$ refers to the field on which the model fit is performed with every field fit independently.}
\label{tab:model}
\end{table*}

The selection function in observed coordinates for field $i$ is $\Prob(\Se_i \mid \mathbf{v}, \Theta_i)$. The stars in the spectroscopic catalogue represent a Poisson realisation of the product of the distribution function and the selection function
\begin{equation}
n_i^\mathrm{spec}(\mathbf{v} \mid \Theta_i) = n_i^{\mathrm{phot}}(\mathbf{v} \mid \Theta_i) \,\times\,\Prob(\Se_i \mid \mathbf{v}, \Theta_i).
\end{equation}
Therefore, we can once more use the Poisson likelihood in Equation~\eqref{eq:poisloglike}, replacing $n^{\mathrm{phot}}(\mathbf{v} \mid \Theta_i)$ with $n^{\mathrm{spec}}(\mathbf{v} \mid \Theta_i)$.

We parameterise the selection function $ \Prob(\Se \mid \mathbf{v}, \Theta_i)$ also as a bivariate GMM in colour-magnitude space with parameters $\widetilde{\boldsymbol{\bep}_i}$ 
\begin{equation}
g_i^{\mathrm{SF}}(\mathbf{v} \mid \widetilde{\bep_i},\Theta_i) = \sum_{k=1}^{\widetilde{K}} \widetilde{w_{i k}}   G(\mathbf{v} \mid \widetilde{\boldsymbol{\mu}_{i k}}, \widetilde{\boldsymbol{\Sigma}_{i k}}),
\end{equation}
The log likelihood is then given by
\begin{equation}
 \begin{split}
 \label{eq:specloglike}
  \ln(\mathcal{L}^\mathrm{spec}(X_i^\mathrm{spec} | &\bep_i, \widetilde{\bep_i})) \propto\\
   -\int \mathrm{d}\mathbf{v}  \,& n_i^\mathrm{phot}(\mathbf{v}\mid\bep_i, \Theta_i)\, g_i^{\mathrm{SF}}(\mathbf{v} \mid \widetilde{\bep_i},\Theta_i) \\
   +& \sum_{j=1}^{N_i^{\mathrm{spec}}} \log \left( n_i^{\mathrm{phot}}(\mathbf{v}_j \mid \bep_i, \Theta_i)\, g_i^{\mathrm{SF}}(\mathbf{v}_j \mid \widetilde{\bep_i},\Theta_i)   \right). \\
  \end{split}
  \end{equation}
  
As the selection function is a probability distribution, it must fall in the range $g_i^{\mathrm{SF}}(\mathbf{v} \mid \widetilde{\bep_i},\Theta_i) \in [0,1]$ for all $\mathbf{v}$. We reparametrise the selection function component weights as $\widetilde{\kappa_{i k}} = \mathrm{logit}\left(\frac{\widetilde{w_{i k}}}{\sqrt{|2\pi\widetilde{\boldsymbol{\Sigma}_{i k}}}}\right)$ with a uniform prior $\widetilde{\kappa_{i k}} \sim \mathrm{U}[-\inf, \inf]$. This constrains the component weights to the range $\widetilde{w_{i k}} \in \left[0, \sqrt{| 2\pi\widetilde{\boldsymbol{\Sigma}_{i k}}|}\right]$ such that the maxima of any Gaussian component is less than or equal to one. 
The sum of selection function Gaussian mixture components must also be less than or equal to one everywhere. Since the maxima of a GMM is non-analytic, we find the roots of the gradient of the GMM using the \textit{hybr} method from {\sc MINPACK-1} \citep{More1980} implemented in {\sc scipy} initialising at the mean of each GMM component. If the value of the GMM at any root is greater than one, the posterior probability is set to zero.

An NIW prior is used for $\widetilde{\boldsymbol{\mu}_{i k}}, \widetilde{\boldsymbol{\Sigma}_{i k}}$ of the selection function.

We fit simultaneously for both the selection function parameters, $\widetilde{\bep_i}$ and photometric density parameters, $\bep_i$ where the prior distributions on $\bep_i$ are the posteriors of the fit to the photometric sample in Section~\ref{sec:dfobs}.

Combining Equation~\eqref{eq:specloglike} with the priors on all parameters the posterior is given by
\begin{equation}
\begin{split}
\ln(\Prob(\bep_i, \widetilde{\bep_i} | X_i^\mathrm{spec})) = \ln(&\mathcal{L}^\mathrm{spec}(X_i^\mathrm{spec} | \bep_i, \widetilde{\bep_i}))\\
- &N + N_i^\mathrm{phot} \ln(N) \\
+ & \mathrm{DIR}(\pi_1, \pi_2...\pi_K | \alpha_1', \alpha_2'...\alpha_K') \\
+ & \sum_{k=1}^K \left[\mathrm{NIW}(\boldsymbol{\mu}_{i k}, \boldsymbol{\Sigma}_{i k} | \mathbf{m}_{0ik}', \lambda_{ik}', \boldsymbol{\Psi}_{ik}', \nu_{ik}') \right]\\ 
+ &  \sum_{k=1}^{\widetilde{K}} \left[\mathrm{NIW}(\widetilde{\boldsymbol{\mu}_{ik}}, \widetilde{\boldsymbol{\Sigma}_{ik}} | \widetilde{\mathbf{m}_0}, \widetilde{\lambda}, \widetilde{\boldsymbol{\Psi}}, \widetilde{\nu})\right].
\end{split}
\end{equation}
To fit the selection function, we need to find $\mathrm{argmin}_{\bep_i, \widetilde{\bep}_i}\left(-\ln\Prob\right)$. All terms in the posterior probability are analytically differentiable so we employ the `Truncated Newton' method in {\sc scipy} which uses the gradient of the probability distribution to converge on the minima. Parameters of the photometric density, $\bep_i$ are initialised at the mean values of the prior distribution.
The selection function parameters are initialised using two methods: 
\begin{itemize}
\item $k$-means clustering with each spectroscopic sample star providing a weighted contribution of  $\frac{1}{n_i^{\mathrm{phot}}(\mathbf{v}_j\mid\bep_i,\Theta_i)}$. The reweighting is akin to approximating the selection function as $n^\mathrm{spec}/n^\mathrm{phot}$.
\item Fit a GMM directly to the spectroscopic data using {\sc BayesianGaussianMixture} with least informative Dirichlet and NIW priors. 
\end{itemize}
The parameters are optimised for both initialisations and the one which leads to the largest posterior probability is used as the best fit. This helps to avoid some local optima in the posterior distribution.

In order to determine the optimal number of Gaussian components to use for the selection function, the algoritm is run for $\widetilde{K} = [1, K]$ components. We do not allow more components than the photometric density GMM as this would likely result in overfitting of the selection function. We use the number of components which minimizes the BIC
\begin{equation}
\mathrm{BIC} =\ln(N_i^\mathrm{spec})d - 2\ln(\mathcal{L}^\mathrm{spec}_\mathrm{max})
\end{equation}

where the number of degrees of freedom $d=(K+\widetilde{K})\times6$ and $\mathcal{L}^\mathrm{spec}_\mathrm{max}$ is the likelihood corresponding to the best fit parameters. To generate a full posterior on the parameters of the photometric density and selection function, we run a set of $6 \times (K + \widetilde{K}) \times 2$ chains for 2000 iterations with \texttt{emcee} \citep{emcee} initialising from a small ball around the best fit parameters.
For every optima of the posterior distribution, there are a set of $\widetilde{K}!$ degenerate optima generated by reordering components. For finding the best fit parameters, it is unimportant which of the optima we sample around, however the posterior on the parameters should not be degenerate. We constrain the means of the photometric density and selection function components to maintain their respective orders in colour and apparent magnitude throughout the \texttt{emcee} iterations.

The models, their parameters and the prior distributions are summarised in Table~\ref{tab:model} for reference. The choice of model hyperparameters is discussed further in Section~\ref{sec:priorhyp}.

\subsection{Intrinsic coordinates}
\label{sec:sfint}

Chemodynamical models of the Milky Way make predictions for the metallicities, masses, and ages of stars at different positions. In order to test these models against observations, the selection function is required to account for the observation biases of the survey. To use the selection function, we require a transformation between observable coordinates (colour and apparent magnitude) and intrinsic coordinates (distance, metallicity, mass, and age). The selection function in terms of the intrinsic coordinates of the stars is
\begin{equation}
\label{eq:specint}
\Prob(\Se \mid  s, \mh, \ms_\mathrm{ini}, \tau, \boldsymbol{\theta}),
\end{equation}
where $s$ is the heliocentric distance, and $\mh$, $\ms_{\mathrm{ini}}$, and $\tau$ are the star's metallicity, initial mass, and age respectively.

Here we follow a similar method for transforming between observable and intrinsic coordinate systems as described in \cite{San15} and \cite{D16a}. Any combination of $\mh$, $\ms_\mathrm{ini}$, and $\tau$ maps to a single set of coordinates, $(c,M)$, which are the colour and absolute magnitude of the star. Any values of $M$ and $s$ uniquely define $m$, the apparent magnitude of the star. Therefore any intrinsic coordinates, $(s, \mh, \ms_{\mathrm{ini}},\tau)$ map to observable coordinates, $(c, m)$ and to a single value of the selection function, $\Prob(\Se \mid \mathbf{v}, \boldsymbol{\theta})$. The selection function in intrinsic coordinates for a star on field $i$ is therefore given by
\begin{equation}
\begin{split}
\Prob(\Se \mid  s, \tau, \mh, \ms_{\mathrm{ini}}, \boldsymbol{\theta} ) = \Prob(\Se  \mid  \mathbf{v}(s, \tau, \mh, \ms_{\mathrm{ini}}), \boldsymbol{\theta}).
\end{split}
\end{equation}
where $\mathbf{v}=(c,m)$. The maps are given by
\begin{equation}
\label{eq:appmag}
m = M + 5\log_{10}\left(\frac{s}{10 \mathrm{pc}}\right)
\end{equation}
and
\begin{equation}
(c, M) = F^{\mathrm{iso}}(\tau, \mh, \ms_{\mathrm{ini}}),
\end{equation}
where $F^{\mathrm{iso}}$ is the mapping introduced by the isochrones.

There are a variety of methods for generating the $F^{\mathrm{iso}}$ map. The most common method is to adopt the nearest isochrone to the values of age and metallicity provided. For this work, we linearly interpolate between isochrones, which improves the accuracy of the transformation.

We use the PARSEC isochrones \citep{PARSEC12} on a grid of $353$ ages and $57$ metallicities. Ages in the range $-2.40\leq \log_{10}\left(\tau/\mathrm{Gyr}\right) \leq 1.12$ with a spacing of $0.01$ dex, and metallicities in the range $-2.192 \leq \mh \leq 0.696$ with a spacing of $0.051$ dex, were considered.

Every isochrone has a maximum initial mass above which a star of the given age and metallicity cannot exist. We generate a scaled initial mass coordinate so that each isochrone varies from $\widetilde{\ms}_{\mathrm{ini}}  \in  [0, 1]$,
\begin{equation}
\widetilde{\ms}_{\mathrm{ini}} = \frac{\ms_{\mathrm{ini}} - \ms_{\mathrm{ini}}^{\mathrm{min}}}{\ms_{\mathrm{ini}}^{\mathrm{max}} - \ms_{\mathrm{ini}}^{\mathrm{min}}},
\end{equation}
where $\ms_{\mathrm{ini}}^{\mathrm{max}}(\tau,\mh)$ is the maximum initial mass value for a star of a given age and metallicity which we determined by linearly interpolating between maximum initial mass of isochrones as a function of $\tau, \mh$. Likewise for $\ms_{\mathrm{ini}}^{\mathrm{min}}(\tau,\mh)$. 

The scaled mass is linearly interpolated along each isochrone. A single set of $\widetilde{\ms}_{\mathrm{ini}}$ values is drawn for all isochrones which samples most heavily where the curvature of the isochrone in colour-apparent magnitude space is greatest. This parameterisation enables us to interpolate colour and absolute magnitude on a regular $\tau - [\mathrm{M/H}] - \widetilde{\ms}_\mathrm{ini}$ grid. The interpolation is the mapping $F^{\mathrm{iso}}(\tau, \mh, \widetilde{\ms}_{\mathrm{ini}}(\ms_{\mathrm{ini}}))$.

%%%%%%%%%%%%%%%%%%%%%%%
\section{Mock tests}
\label{sec:mock}

\begin{figure}
\includegraphics[width = \columnwidth]{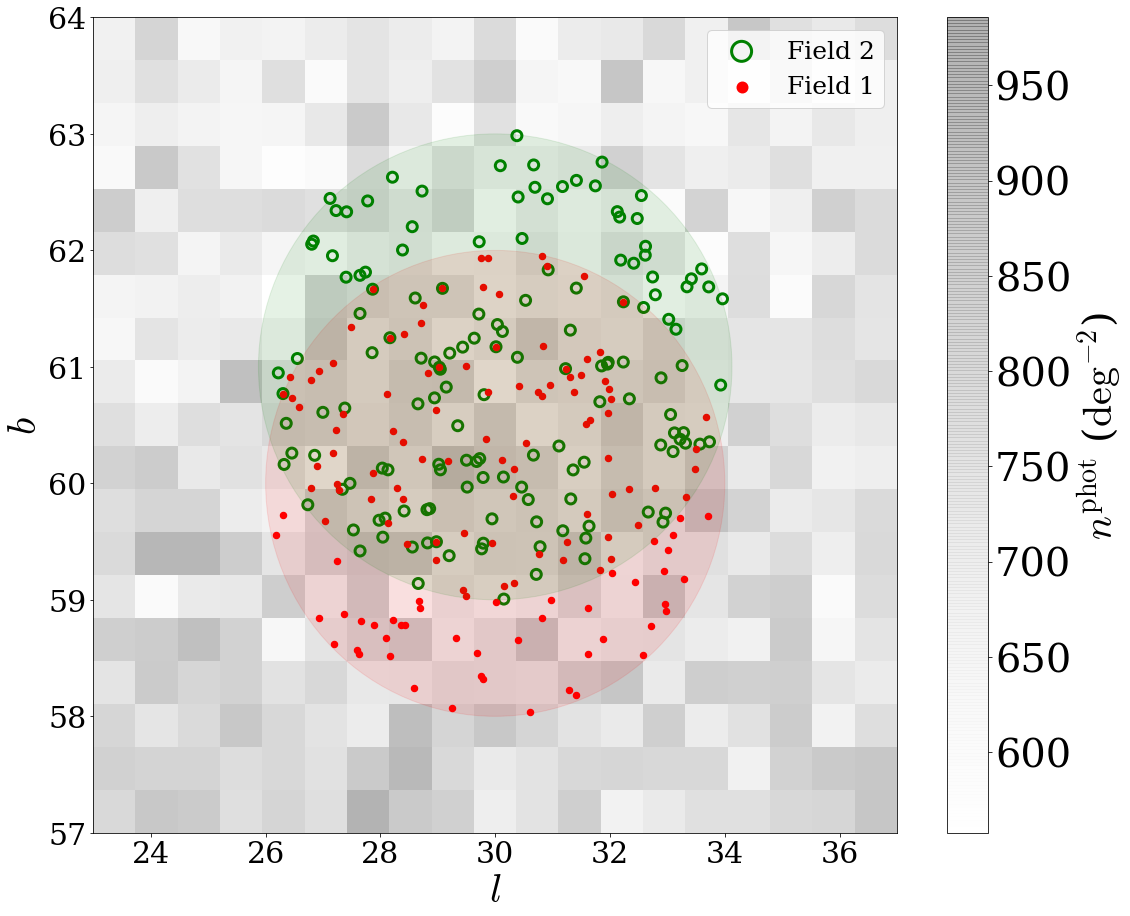}
\caption{The mock \texttt{Galaxia} model uses two fields offset by $1\deg$ latitude represented here by the red and green shaded regions. The grey scale bins give the number density of stars in the photometric sample which systematically decreases further from the plane. Applying the Flat selection function returns the stars represented by red points and green circles for fields 1 and 2 respectively. In the region of overlap there are cases where stars are selected on both fields.}
\label{fig:galfields}
\end{figure}
\begin{figure*}
\includegraphics[width = \textwidth]{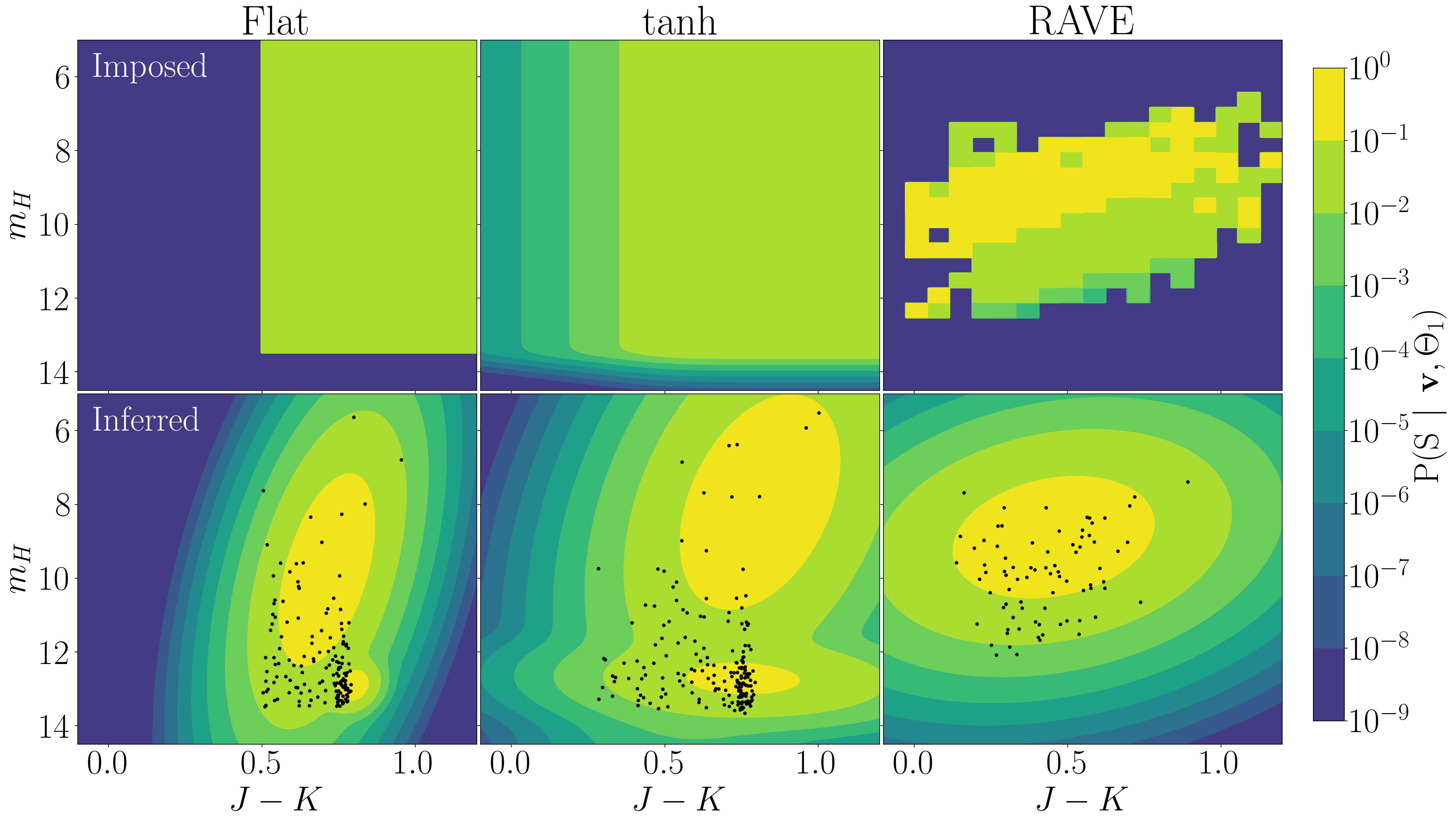}
\caption{We apply three different selection function models to the \texttt{Galaxia} mock sample, Flat (left), tanh (middle) and RAVE (right). Top: The models are defined as a function of colour and apparent magnitude. Bottom: Using the BIC as our model selection criteria, two GMM components are fit for each model shown by the contours. The model broadly recovers the form of the imposed selection function in regions occupied by the selected samples which are shown as black points.}
\label{fig:appliedsf}
\end{figure*}

To test the performance of the method presented in Section~\ref{sec:method}, we apply it to mock samples with known selection functions. 
%We generate a mock catalogue of stars and impose a \textit{known} selection function. We calculate  $\Prob(\Se \mid c, m, \bth)$ (Section~\ref{sec:sfobs}) and $\Prob(\Se \mid s,\mh,\ms_\mathrm{ini},\tau,\bth)$ (Section~\ref{sec:sfint}) and compare the derived selection function to that imposed.

%-------------------------------------------------
\subsection{\texttt{Galaxia} mock catalogue}

\begin{figure*}
\includegraphics[width=\textwidth]{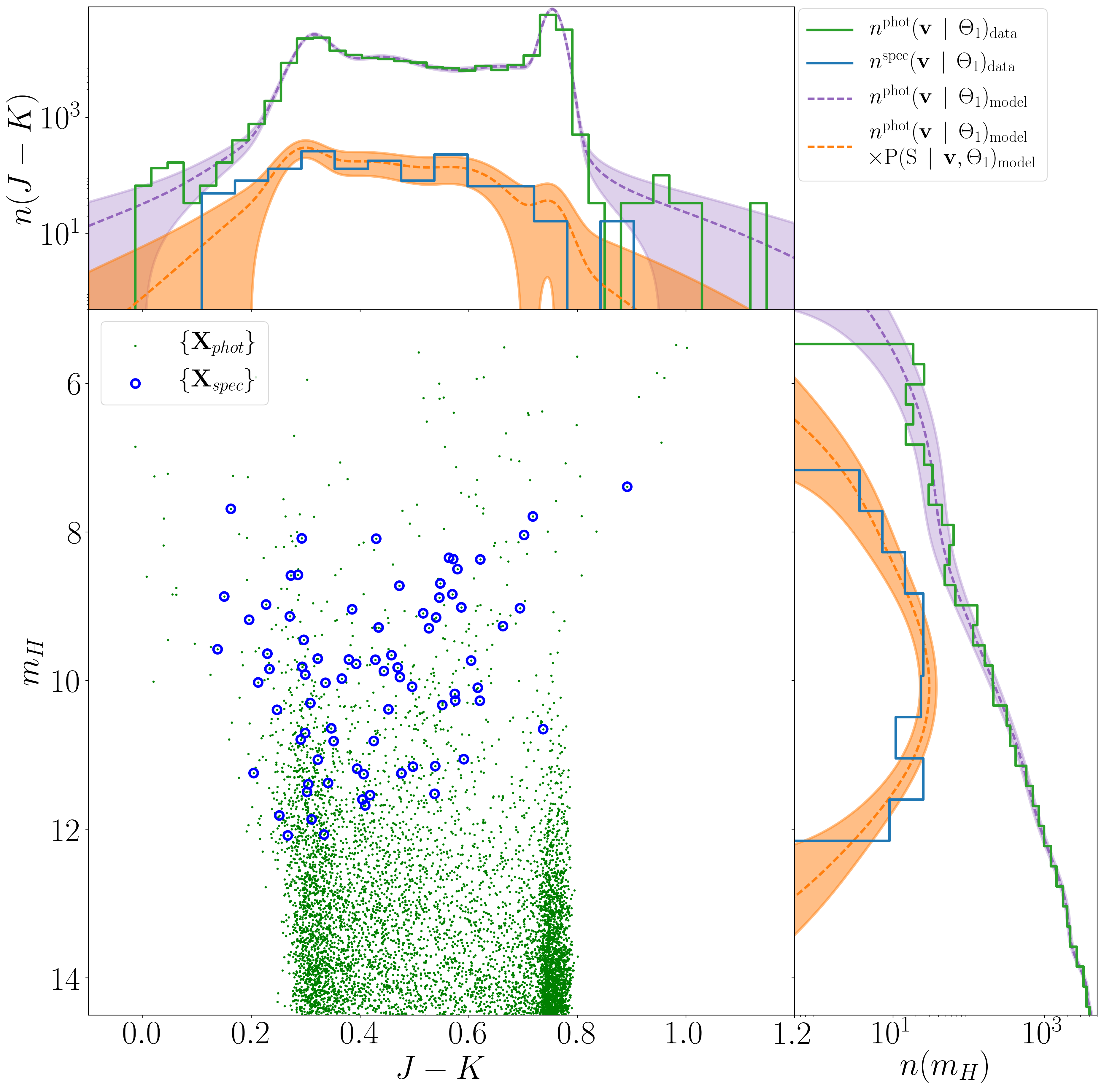}
\caption{The selection function selects a subset (blue circles) of the photometric sample (green points) to be included in the spectroscopic catalogue where in this case we are showing the RAVE mock applied to field 1. The model fits 14 GMM components to the photometric sample for which purple dashed lines show the 1D projection. The photometric sample distribution (green solid histograms) mostly fall within Poisson noise uncertainty (purple shaded region) of the GMM fit. The selection function is fit with 2 GMM components such that the spectroscopic sample is modelled by the 28 component product of the two GMMs, represented by the orange dashed line. The spectroscopic subsample (blue solid histogram) largely falls within the Poisson noise uncertainty (orange shaded region) of the model fit.}
\label{fig:cmscatterhist}
\end{figure*}
\begin{figure*}
\includegraphics[width=\textwidth]{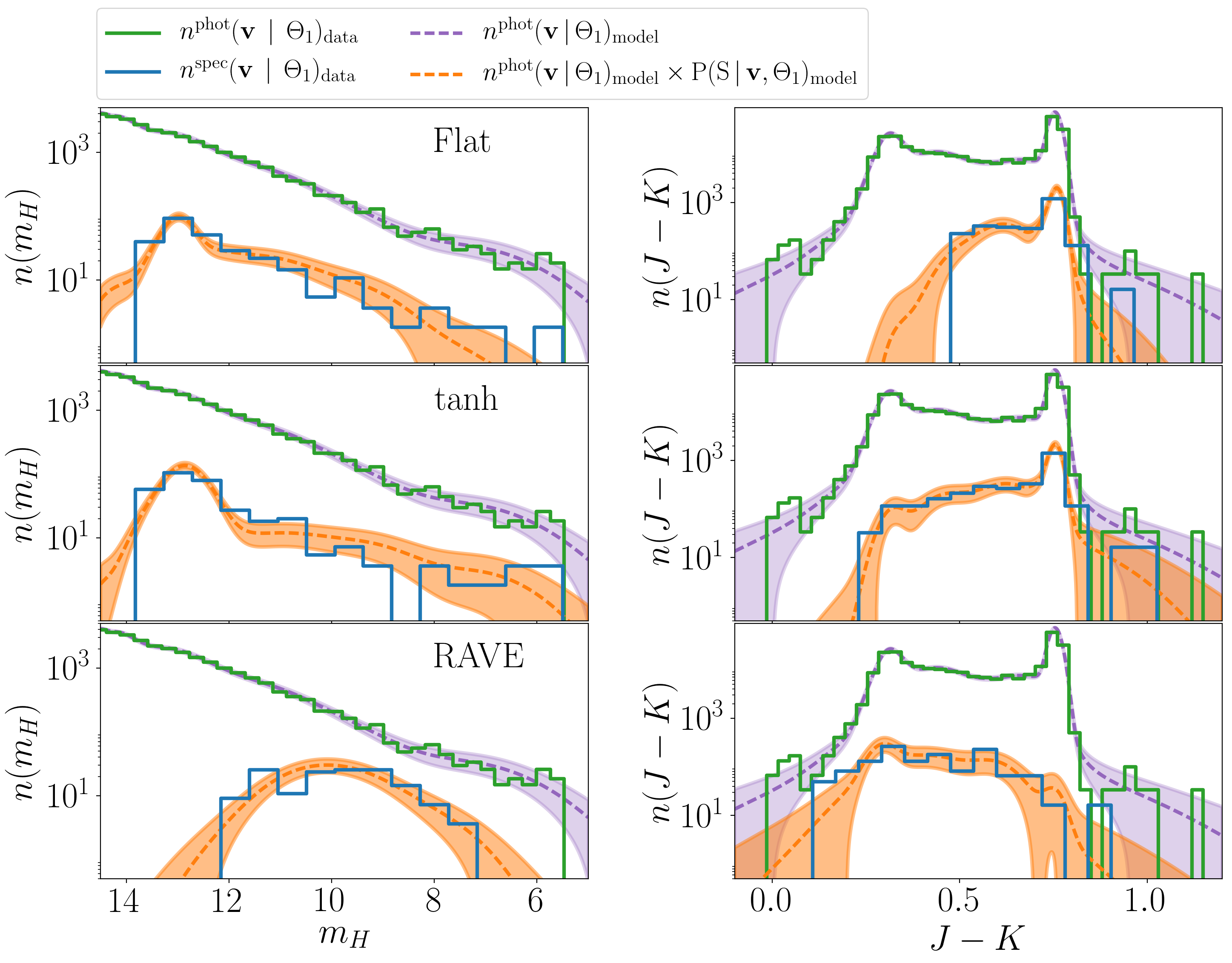}
\caption{The same as the side panels of Figure~\ref{fig:cmscatterhist} for the Flat (top), tanh (middle) and RAVE (bottom) models applied to field 1 of the \texttt{Galaxia} mock. In the majority of cases the spectroscopic model, given by the product of the photometric sample and selection function GMMs, traces the spectroscopic samples within Poisson noise uncertainty. The noticeable limitation is in the Flat model (top panels) where the smooth GMM cannot reproduce the sharp cut-off in colour ($J-K=0.5$) and apparent magnitude ($m_H=13.5$).}
\label{fig:mhist}
\end{figure*}

We generate our mock catalogue using the \texttt{Galaxia} code \citep{Galaxia11}. The thin disk is generated from the analytic Besan\c con model \citep{Bes03}. Non-circular motions in the plane of the disk are introduced through the \citet{Shu69} DF, and the \citet{Bullock05} N-body models simulate any substructure in the halo. The code synthesises a population of stars with coordinates, trajectories and intrinsic properties including age, metallicity and mass. Given the age, metallicity, initial mass and distance, \texttt{Galaxia} also calculates the colour and apparent magnitude of stars as observed from the Sun by employing the nearest Padova isochrone \citep{Marigo08} from a grid of 182 ages and 34 metallicities. The PARSEC \citep{PARSEC12} isochrones however present a significant update on the Padova isochrones in terms of revisions to major input physics such as the equation of state, opacities, nuclear reaction rates, and inclusion of the pre-main sequence phase. The nearest isochrone calculation method is also less accurate than the interpolation method we employ with PARSEC isochrones discussed in Section~\ref{sec:sfint}. For these reasons, we recalculate the \texttt{Galaxia} apparent magnitudes and colours using our method.

We sample from \texttt{Galaxia} with an $H$-band magnitude limit, $m_\mathrm{H}<15$, similar to the limitations of the 2MASS survey \citep{Sk6}. The $H$-band magnitude limit was placed using \texttt{Galaxia} such that stars were filtered out based on the built-in Padova magnitudes. The difference between this and our own magnitude calculation leads to some stars dimmer than the magnitude cut being included and some brighter stars being excluded. This does not significantly affect our tests. 

We will refer to this catalogue as the photometric sample.

%----- FIGURES
\begin{figure*}
\includegraphics[width=0.9\textwidth]{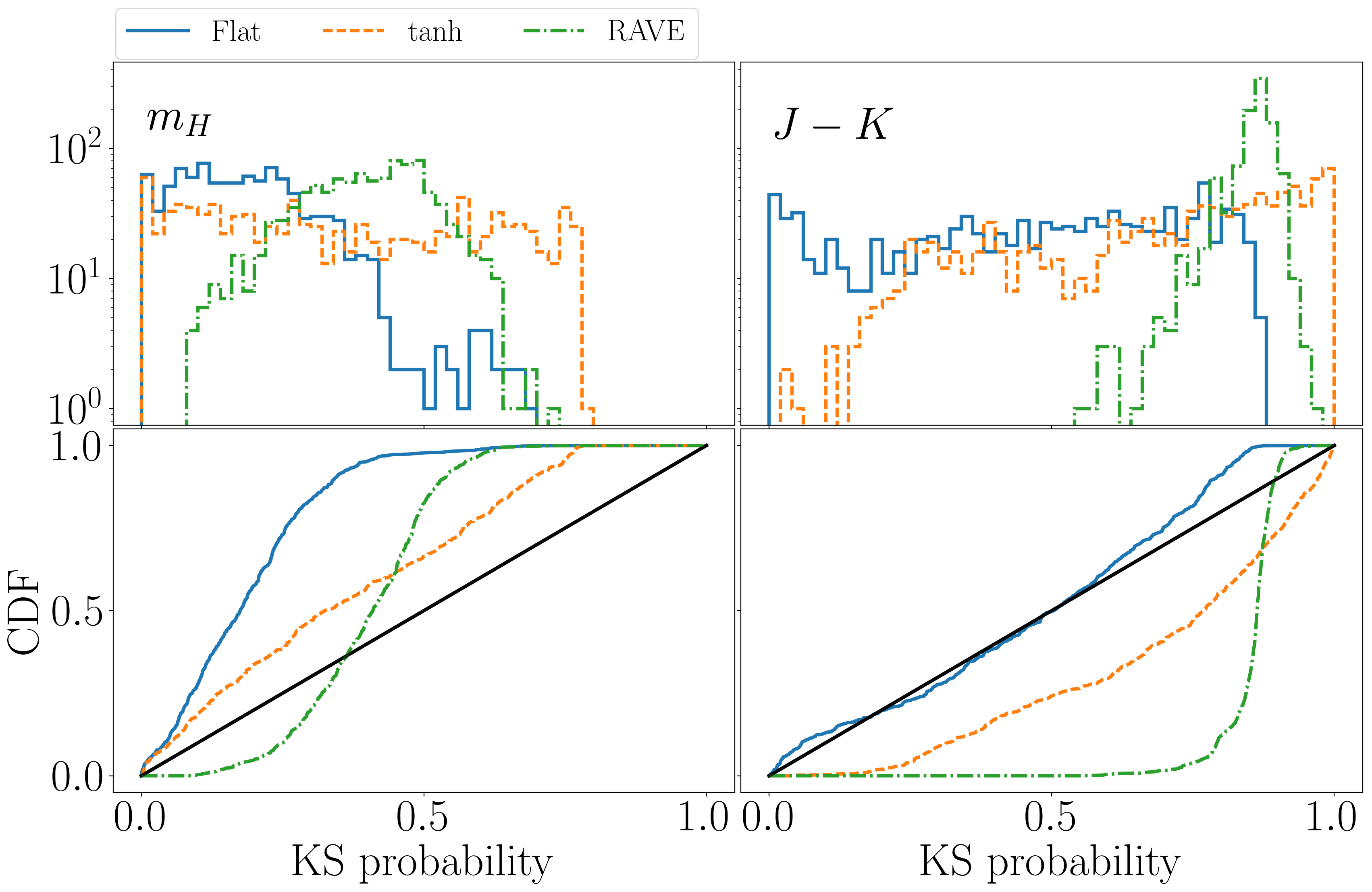}
\caption{The 1D one sample KS probability is evaluated for the field 1 spectroscopic sample against the GMMs of 1000 random draws from the \texttt{emcee} chains. In apparent magnitude (left) the Flat selection function shows underfitting (blue solid) as the distribution of probabilities (top) is biased towards 0 and the CDF (bottom) sits well above a uniform distribution (black solid). In colour (right) the RAVE sample has overfit the data (green dot-dashed) as the KS probabilities are biased towards 1. The remainder of the tests are closer to uniformly distributed and demonstrate reasonable fits to the data against the given coordinates.}
\label{fig:ks2}
\end{figure*}
\begin{figure}
\includegraphics[width=\columnwidth]{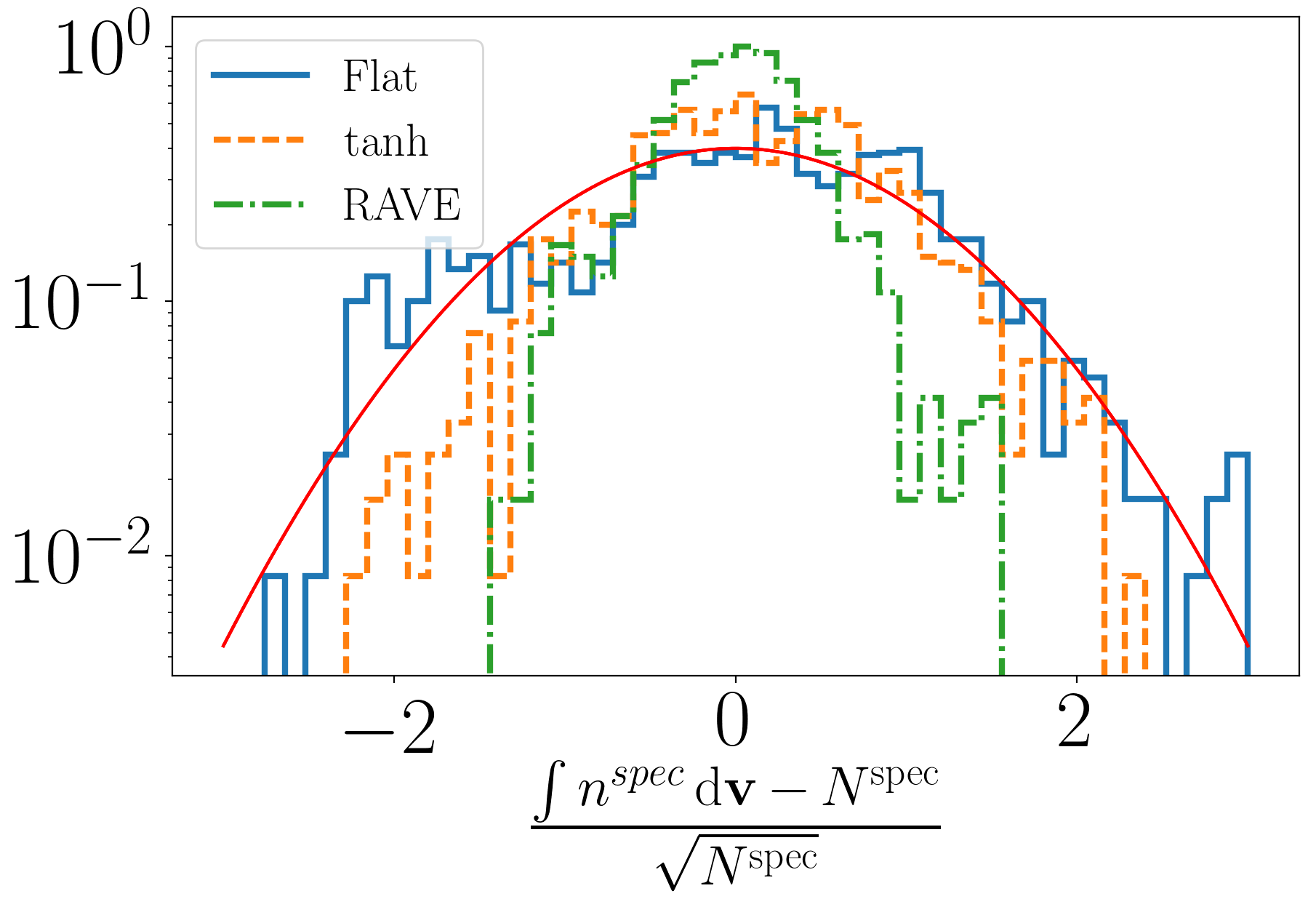}
\caption{The integrals over the GMM fits are compared to the number of stars in the spectrograph subsample. A well-fit model would be normally distributed around \mbox{$\int n^\mathrm{spec}\mathrm{d}\mathbf{v} = \sqrt{N^\mathrm{spec}}$} with dispersion $\sqrt{N^\mathrm{spec}}$ demonstrated by the red solid line. The Flat model fit (blue solid) accomplishes this whilst the tanh (orange dashed) and RAVE (green dot-dashed) fits show a tighter distribution indicative of overfitting.}
\label{fig:integral_1field}
\end{figure}
\begin{figure*}
\includegraphics[width=0.9\textwidth]{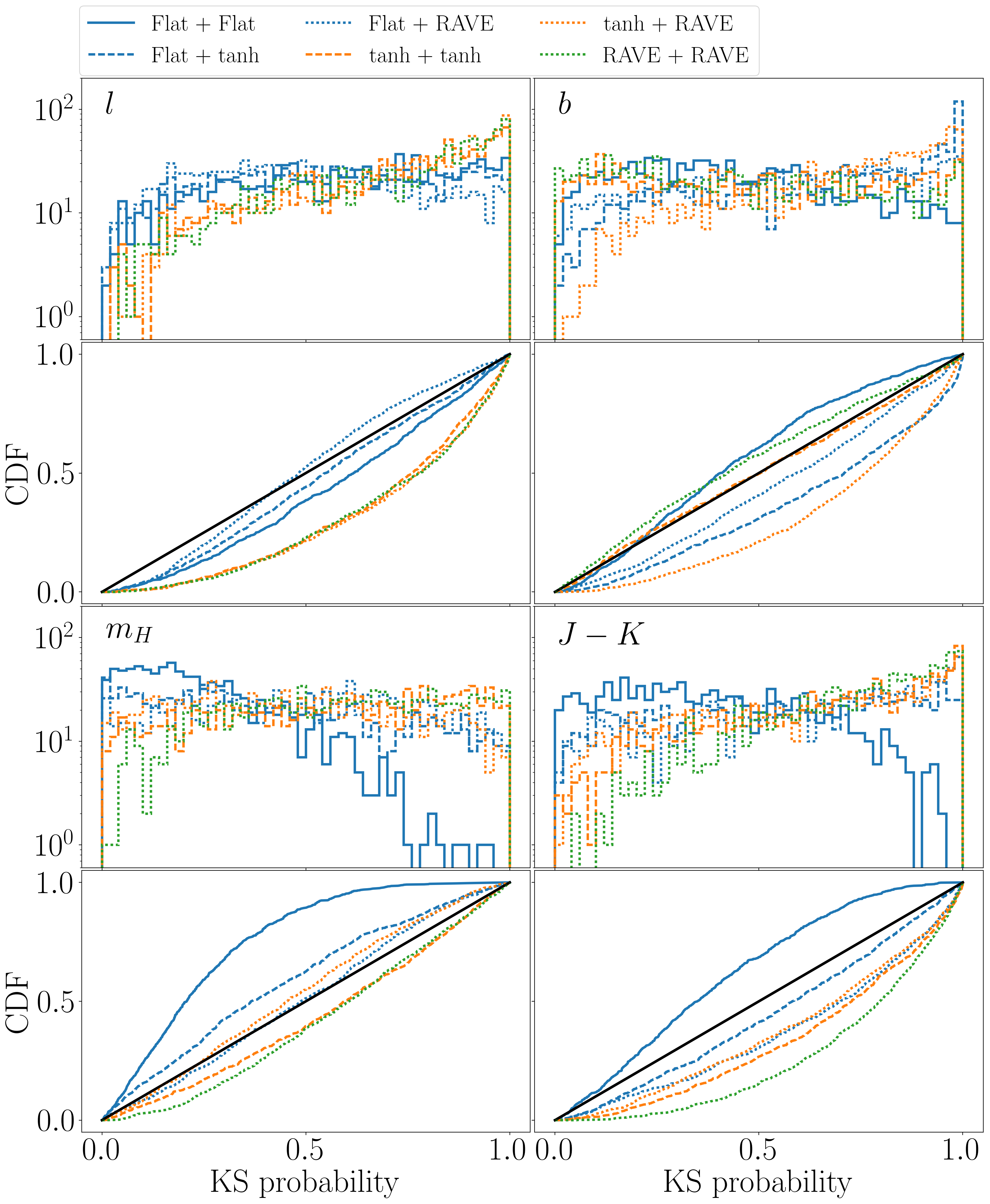}
\caption{Two sample KS tests are used for fields 1 and 2 combined as the distributions are non-analytic. 1000 sets of selection function parameters are drawn from the \texttt{emcee} chains and applied to generate subsamples of the photometric catalogue. Six combinations of the three selection functions are used for the two fields and most show close to uniform probability distributions with some weak overfitting against galactic longitude (top left) and latitude (top right). The most significant deviation is the Flat+Flat model (blue solid) which is underfit in apparent magnitude (bottom left) and colour (bottom right).}
\label{fig:ks2match}
\end{figure*}
\begin{figure}
\includegraphics[width=\columnwidth]{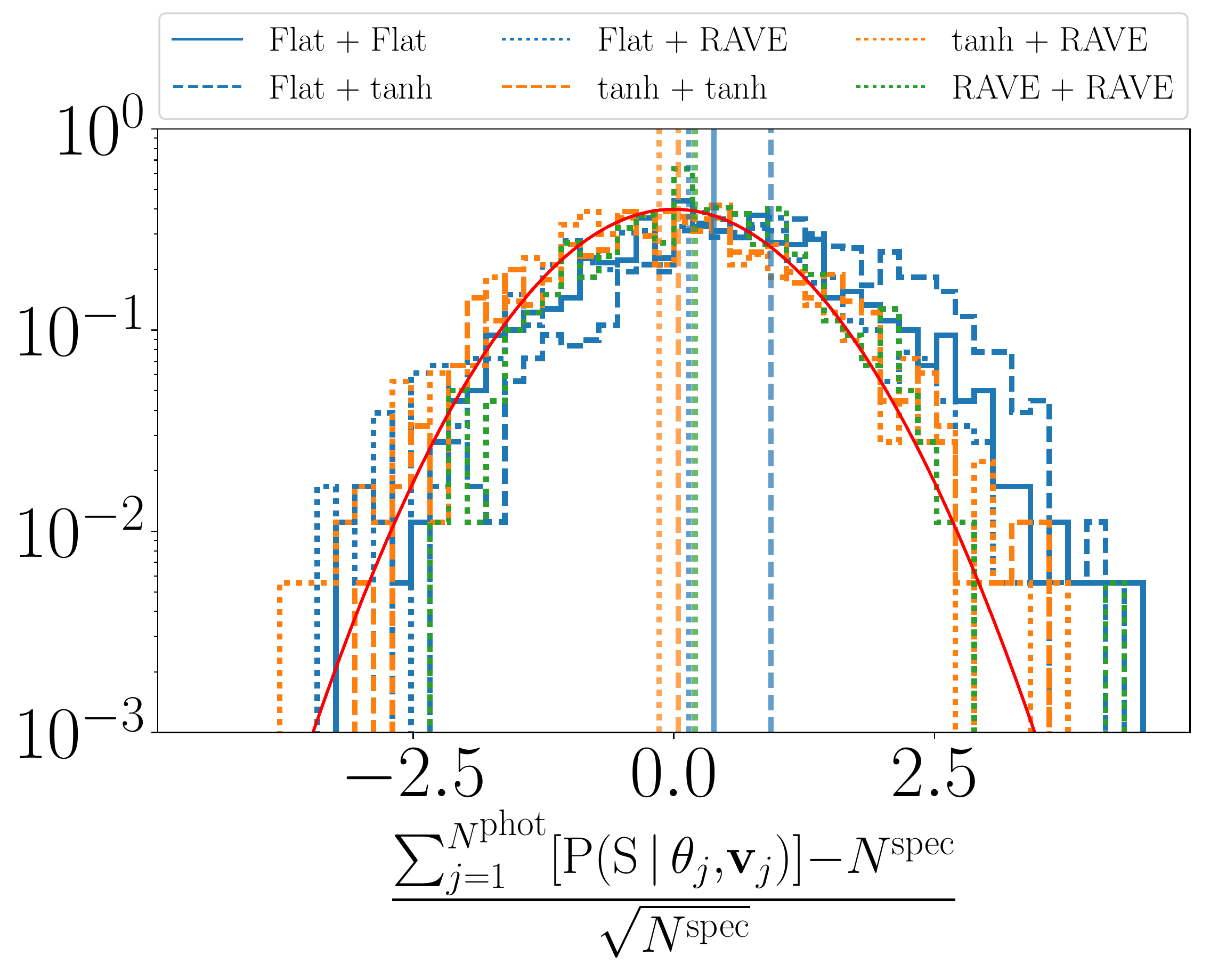}
\caption{By taking the sum of probability of selection of all stars in the photometric sample for a given selection function, we obtain an estimate of the integral over the spectroscopic distribution. All models are normally distributed with widths similar to that expected by Poisson noise of the spectrograph sample and the means of the distribution (vertical lines) are correctly centered on zero. The red solid line is a normal distribution with zero mean unit variance for comparison. Only the Flat+tanh model (blue dashed) systematically overestimates the normalisation by approximately one standard deviation.}
\label{fig:integral_2field}
\end{figure}

\begin{figure*}
\includegraphics[width=\textwidth]{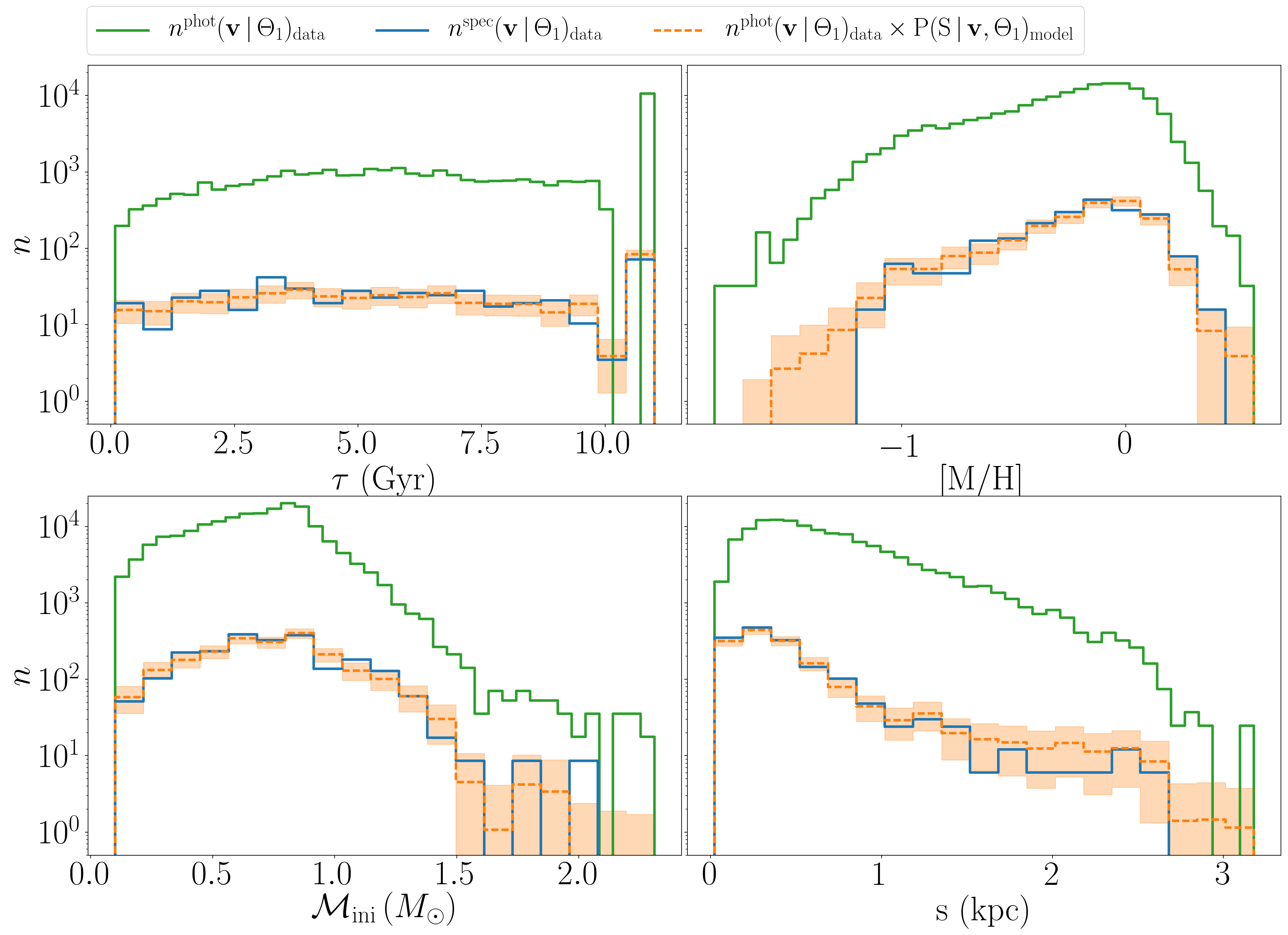}
\caption{Similar to Figure~\ref{fig:mhist} now for intrinsic coordinates, the distributions of stars in the photometric sample (green solid) and spectroscopic sample (blue solid) compared to the sum of selection function probabilities of the photometric sample in bins (orange dashed). For this example, field 1 and 2 have been used with tanh and RAVE selection functions applied respectively. For the majority of bins the spectroscopic sample falls within Poisson noise uncertainty (orange shaded) of the model fit.}
\label{fig:histintrinsic}
\end{figure*}
\begin{figure}
\includegraphics[width=\columnwidth]{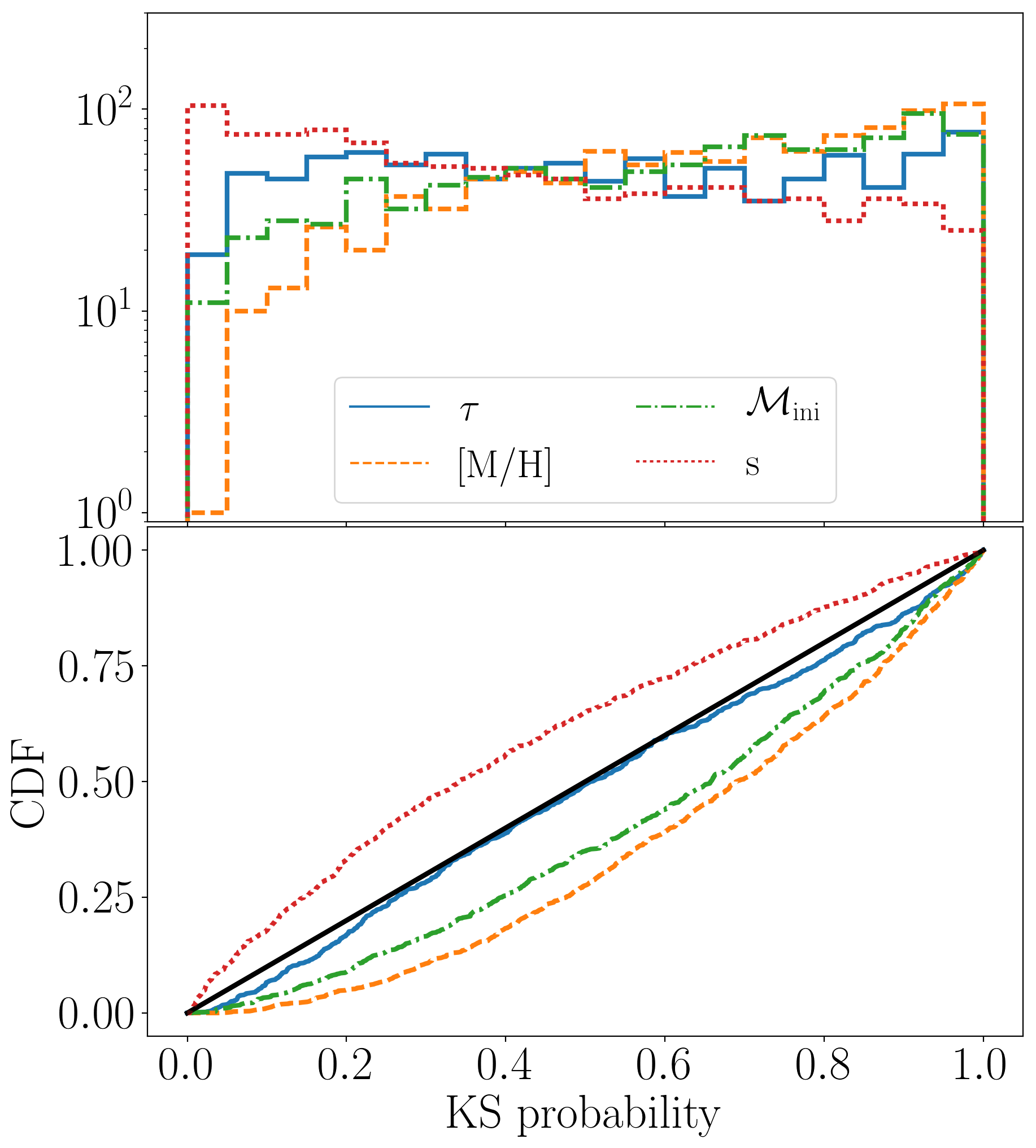}
\caption{As in Figure~\ref{fig:ks2match}, the two sample KS test is applied but here as a function of each intrinsic coordinate with the tanh and RAVE selection functions applied for fields 1 and 2. Much of the over and underfitting seen in Figure~\ref{fig:ks2match} is averaged out in these coordinates and the models demonstrate extremely good fits to the data.}
\label{fig:KSintrinsic}
\end{figure}

%-----------------------------------------
\subsection{Imposed selection function}

The selection function is applied across two fields with coordinates
\begin{equation}
\label{eq:galaxiafps}
l, b =
\begin{cases}
30, 60 \quad (\text{Field 1}),\\
30, 61 \quad (\text{Field 2}).\\
\end{cases}
\end{equation}
Each field has a half-opening angle of $2\mathrm{deg}$ and hence a solid angle of $12.6\mathrm{deg}^2$. The locations of the fields in galactic coordinates are given in Figure~\ref{fig:galfields} where the red and green shaded regions are fields 1 and 2 respectively and the greyscale background shows the number density of stars in the photometric sample.

For each field we apply three different selection functions:
\begin{itemize}
\item Flat: Selection boundaries of $m_H < 13.5$ and $J-K > 0.5$. The value of the selection function is $0.1$ within the boundaries and $0$ outside.
\item tanh:
    \begin{equation}
        \begin{split}
    \Prob(S|m_H, J-K) = 0.1 &\times \left(1- \tanh\left(\frac{m_H - 13.5}{e^{-2}}\right)\right)/2 \\
                        & \times \left(1- \tanh\left(\frac{J-K - 0.5}{e^{-2}}\right)\right)/2.
    \end{split}
    \end{equation}
This is of a similar form to the selection function proposed for SEGUE G-dwarfs by \citet{Bovy12}.
\item RAVE: We take 10 randomly selected fields from the RAVE survey \citep{K13} along with any 2MASS  \citep{Sk6} stars on the same fields, bin the stars in $m_{H, \mathrm{2MASS}} - (J-K)_\mathrm{2MASS}$ and use the ratio of RAVE to 2MASS stars in each bin as the selection function probability.
\end{itemize}
The three applied selection functions are plotted in top panels of Figure~\ref{fig:appliedsf}. The red and green scatter points in Figure~\ref{fig:galfields} are the stars selected by the Flat selection function to demonstrate the setup.

In the region of overlap of the fields, stars may be selected by either or both of the fields. We include every star selected by at least one field however we also record which fields each star was selected by as this is important for calculating the selection function. For instance, when deriving the selection function for field 1, we must use all stars which were selected by field 1 even if they were also selected by field 2. Likewise for field 2, we must include all stars selected by field 2 even if they were also selected by field 1. The effective double counting of selection functions in overlapping regions is accounted for by the union calculation as described in Section~\ref{sec:defpatch}.

%-----------------------
\subsection{Results}
\label{sec:mockresults}

The \texttt{Galaxia} photometric and spectroscopic catalogues are used to calculate the selection functions in observable coordinates, ${\Prob(\Se \mid \bth, c, m)}$, and intrinsic coordinates, \mbox{$\Prob(\Se \mid \bth, s, \mh, \mathcal{M}_\mathrm{ini},\tau)$} following the methods described in Section~\ref{sec:method}. 

We first discuss the values of hyperparameters used for the GMM priors. We will present and test the posterior for field 1 as a function of colour and apparent magnitude. We then show the fit to both fields in observable coordinates and position on the sky which tests how the model handles overlapping fields. Finally we present the selection function in intrinsic coordinates.

%-------------------
\subsubsection{Prior hyperparameters}
\label{sec:priorhyp}

For the prior distributions on the photometric density and selection function we require hyperparameter values which encode our prior knowledge about the distribution. We assume that we know all data falls within the apparent magnitude and colour ranges $m_H \in [4, 15]$ and $J-K \in [-0.1, 1.2]$.

For the NIW hyperparameters on the photometric density fit we use {$\mathbf{m}_0 = \mathrm{med}\left[\mathbf{v}_\mathrm{min}, \mathbf{v}_\mathrm{max}\right] = (9.5, 0.55)$} which is at the center of our prior range. We set $\lambda = 10^{-4}$. By choosing a small value for $\lambda$ we ensure that the prior is extremely uninformative on the mean. For the covariance we choose {$\boldsymbol{\Psi} = \mathrm{diag}((\frac{\mathbf{v}_\mathrm{max}-\mathbf{v}_\mathrm{min}}{2})/5)^2) = \mathrm{diag}(1.21,  0.0169)$} such that the standard deviation is $1/5$ the prior range in $m_H$, $J-K$. The precise choice of $1/5$ is somewhat arbitrary and chosen as it leads to reasonable numbers of Gaussian components in the photometric density and selection function GMMs in all cases. The degrees of freedom is chosen as $\nu = 2$ which is the most uninformative choice. For the Dirichlet prior we choose a concentration prior of $\alpha_k = 1/K$ for all components which is the least informative prior.

We use the same hyperparameters for the NIW prior on the selection function as those used for the photometric density fit. This is because we have no more information about the selection function than we did about the photometric density other than that it is only important in the given colour and magnitude range. Having found the best fit parameters, we also run a set of \texttt{emcee} chains as discussed in Section~\ref{sec:sfobs} to generate posterior samples on all of the photometric density and selection function parameters.

The choices of hyperparameter values are summarised in the right-hand column of Table~\ref{tab:model}.

 %-------------------------------
\subsubsection{Observable coordinates}

In this section we demonstrate the selection function fit to field 1 of the mock sample in colour-apparent magnitude space. To minimize the BIC, the optimal number of photometric density GMM components was 14 whilst the Flat, tanh and RAVE selection functions were fit with 2, 2 and 1 components.

For the RAVE mock, the distribution of photometric and spectroscopic points and the GMM distribution fits are given in Figure~\ref{fig:cmscatterhist}. For a successful fit we would expect the centers of $68\%$ of histogram bins to fall within the Poisson uncertainties of the model and we see that this appears to be the case for our fit to the RAVE selection function against both colour and apparent magnitude. The same number density distributions for all three mock selection functions are shown in Figure~\ref{fig:mhist}. For the Flat model (top panel) the smooth GMM selection function fails to correctly fit the sharp cuts at $m_H=13.5$ and $J-K=0.5$. This is a predictable limitation of attempting to fit a discontinuous model with a smooth function. In all other cases the histograms of the spectroscopic sample mostly fall within the Poisson noise uncertainties of the products between our fits to the photometric density and selection function. This provides a qualitative demonstration that the method is providing reasonable results.

For a more quantitative assessment we use the Kolmogorov-Smirnov (KS) test. We randomly draw 1000 samples from the last 500 iterations of each \texttt{emcee} fit to the spectroscopic sample. The KS probability (p-value of the KS test) is the probability that the spectroscopic sample is consistent with being drawn from the given probability distribution, in this case the product of the photometric and selection function GMMs which is itself a GMM with $K\times\widetilde{K}$ components.

 The KS statistics are computed in 1D as a function of colour and apparent magnitude separately. For a good fit to the data, the KS probabilities will be uniformly distributed in the range $[0,1]$. If the probability is skewed towards $0$, the method has underfit or failed to fit the data. If the probability is skewed towards $1$, the method has overfit the data and hence the selection function would not generalise well to a new dataset. The most extreme example of this would be placing delta functions on every star which would achieve a KS probability of 1.

We expect that the KS statistic should demonstrate some overfitting since the test does not consider the prior information provided by the photometric sample.

Histograms and cumulative distributions of the KS tests for field 1 against colour and apparent magnitude are shown in Figure~\ref{fig:ks2}. Due to the sharp cut off in the Flat model, the KS probabilities demonstrate that the GMM underfits the selection function as a function of magnitude. Curiously, the same problem is not seen as a function of colour where the Flat model is almost perfectly fit. KS tests are only weakly sensitive to the wings of the distribution and we postulate that this is why the sharp cutoff issue is not picked up in colour space. The RAVE and tanh models provide reasonable fits as a function of apparent magnitude however the RAVE model is heavily overfit as a function of colour. The cause of this overfitting is unclear.

To test the normalisation of the photometric density and selection function we compare the integral over colour and apparent magnitude of the GMMs with the number of objects in the spectroscopic sample. For large samples, we would expect the difference between the integral and true count to be Gaussian distributed with variance equal to the size of the sample. Figure~\ref{fig:integral_1field} shows the distribution of these integrals from the 1000 \texttt{emcee} parameter draws re-centered and re-scaled by the spectroscopic sample number counts. All distributions are centered around zero which implies that the fits are well normalised. The fits to the RAVE model show a tight distribution which is another indication of overfitting to the spectroscopic data.

 %-----------------------------
\subsubsection{Overlapping fields}
\label{sec:overlaptest}

On field 2 the photometric density is fitted with 14 GMM components and the Flat, tanh and RAVE selection functions with 2, 3 and 1 components respectively to minimize the BIC. We will now only consider the selection function resulting from the combination of fields 1 and 2.

The selection function is now a superposition of two GMMs as a function of galactic coordinates. As such the KS one-sample statistic can no longer be used as we don't have an analytic distribution to fit. Instead we draw random samples from the photometric catalogue with the probability of inclusion given by the model selection function as a function of galactic coordinates, colour and apparent magnitude. We compute the two-sample KS probability against the spectroscopic catalogue which gives the probability that samples are drawn from the same probability distribution.

In Figure~\ref{fig:ks2match} we show the distribution of KS probabilities for fields 1 and 2 with all combinations of the Flat, tanh and RAVE selection functions applied. The Flat+Flat model (blue solid line) systematically underfits the data in both apparent magnitude and colour which, as discussed earlier, is caused by the sharp cuts in the selection function which a smooth GMM cannot correctly reproduce. All other combinations of fields produce close to uniform distributions of KS probabilities with weak under and overfitting against different coordinates. 

To test the normalisation of the selection function we compare the sum of the selection function over the stars in the photometric sample with the number in the spectroscopic sample. This sum is the mean sample size which would be produced by drawing many selection function weighted samples from the photometric catalogue. The re-centered and re-scaled distributions for all field combinations are shown in Figure~\ref{fig:integral_2field}. We see significantly less overfitting here with all distributions reproducing the expected Poisson noise. Only the Flat+tanh model (blue dashed) systematically overestimates the normalisation however the offset is only at the one standard deviation level so we don't consider this to be a significant issue.

These tests demonstrate the effectiveness of the union method described in Section~\ref{sec:defpatch} for evaluating the selection function of overlapping fields of the spectroscopic survey. This also shows that the method provides good fits to selection functions with very different properties. A caveat is that the GMM is not perfectly suited to fitting selection functions with discontinuous changes however even in these cases reasonable fits can still be achieved.

We also show here the power of our method for calculating selection functions of combined surveys. For example we could apply this method to generate a single selection function for the combined APOGEE and RAVE catalogues given their individual selection functions.

 %--------------------------------
\subsubsection{Intrinsic coordinates}

Our final test is on selection function as a function of intrinsic coordinates (age, metallicity, mass and distance) using the mapping laid out in Section~\ref{sec:sfint}. We test the selection function for the two field sample with the tanh and RAVE selection functions applied to fields 1 and 2 respectively. A histogram for the model fit is generated by taking the binned sum of the selection function probabilities of the photometric sample and is shown by the orange dashed histogram in Figure~\ref{fig:histintrinsic}. The blue solid line shows the histogram for the spectroscopic sample and this falls within Poisson noise uncertainties of the model in the majority of bins.

Similar to Section~\ref{sec:overlaptest}, we generate 1000 mock samples from the photometric catalogue weighted by the inferred selection function and calculate the goodness of fit from the two-sample KS statistic. The distribution of KS probabilities is given in Figure~\ref{fig:KSintrinsic}. As we saw in observable coordiantes, the distribution of KS probabilities is near uniform with very weak under fitting against distance and overfitting in initial mass and metallicity. This demonstrates that our method is also well suited to determining selection functions in intrinsic coordinates.

%%%%%%%%%%%
\section{Discussion}
\label{sec:discussion}

In this paper, we have introduced a novel method for deriving selection functions of spectroscopic surveys. We have also demonstrated the success of this method on a set of test cases using \texttt{Galaxia}. In this section, we discuss some key points to consider when applying the method, and detail potential improvements to come.

\subsection{Choice of photometric catalogue}
When calculating the selection function of any spectroscopic survey, a photometric survey needs to be specified which may be assumed complete in the region of the colour-apparent magnitude space explored. The choice of photometric catalogue is dependent on the characteristics of the survey. The photometric catalogue should cover the whole footprint of the spectrograph or else a combination of photometric catalogues should be used. It is beneficial to use a photometric catalogue with observing bands closely matching the spectrograph wavelength range as this enables prior information on the spectrograph's selection limitations to be applied more easily.

That said, particularly for low-latitude observations, dust attenuation is a significant factor, which suggests that infrared photometric surveys may be more appropriate. We will discuss the inclusion of dust attenuation to the intrinsic selection function in Section~\ref{sec:dust}.

Gaia DR2 \citep{GaiaDR2} represents the largest survey of the Milky Way to date and will be a complete photometric survey for the magnitude ranges of many spectrographs, particularly in high latitude fields. The selection of Gaia DR2 as a function of $l, b, m_G$ will be provided by Boubert \& Everall (2020a, in prep) and can be used to test whether it is complete in the required magnitude range.

\subsection{Error convolution}

The selection function we have derived here is the probability of selection given measured properties of the stars. By convolving the likelihood with the measurement uncertainty of the photometries of each star, we can in principle derive the selection function given true properties of the stars.
By virtue of defining the selection function as a GMM in colour-apparent magnitude space the convolution is analytic and as such the calculation is computationally feasible.

We do not present this here, but consider it as a potential avenue to pursue in the future.

\subsection{Dust attenuation}
\label{sec:dust}
In this work we have not considered the impact of interstellar dust on the derived selection function. The observable selection function is dependent on dust attenuation only through it's effect on colour and apparent magnitude. Our inferred selection function given colour and apparent magnitude will be unchanged. 

Dust attenuation changes the mapping from intrinsic coordinates to observable coordinates discussed in Section~\ref{sec:sfint}. The effect of dust on the transformation from absolute to apparent magnitude of the star is given by
\begin{equation}
\label{eq:appmagext}
m_x = M_x + 5\log_{10}\left(\frac{s}{10 \mathrm{pc}}\right) + A_x(l, b, s),
\end{equation}
where $x$ represents the observation band being used for apparent magnitude in the selection function. Likewise the colour will also need to be corrected for dust reddening in the mapping from intrinsic coordinates to observables.

This will be an especially important consideration for low-latitude fields. To do this, we require an adequate 3D all-sky dust map. 

\citet{BovyDust16} construct a composite map that patches together maps of \citet{Marshall06}, \citet{Sale14} and \citet{Green15}. A significant improvement on this is provided by \citet{Green19} for $\mathrm{dec}>-30\mathrm{deg}$ which is extrapolated to the whole sky using Gaia photometries in Boubert \& Everall (2020b, in prep).

We can include these maps in our model as an extinction and reddening term in the intrinsic to observable coordinate mapping described in Section~\ref{sec:sfint}.

\subsection{Gaia RVS}

We have not considered the Gaia Radial Velocity Spectrograph \citep[RVS][]{GaiaDR2RVS} in this paper as the strong dependence on the Gaia scanning law means that our method is not the best for this catalogue. In Boubert \& Everall (2020b, in prep) we develop an independent method for evaluating the selection function of Gaia RVS which properly treats the intricate behaviour of the scanning law. These works will prove complimentary to one another for studies of the chemo-dynamic properties of the Milky Way.

\subsection{Future work}

We will be applying this method to a series of currently available multifibre spectroscopic datasets including APOGEE DR14 \citep{abolfathi+18}, LAMOST DR5 \citep{LAMOST12}, RAVE DR5 \citep{K17}, Gaia-ESO DR3 \citep{GaiaESO12}, GALAH DR2 (\citet{Galah2}) and SEGUE (\citealp{SEGUE09}) and their future data releases. Further into the future, we will apply this algorithm to the first data releases of upcoming multifibre spectrographs, WEAVE \citep{WEAVE12}, MOONS \citep{MOONS14} and 4MOST \citep{4MOST12}.

All new selection functions will be made publicly available on the {\sc Python} module \textsc{selectionfunctions} (\url{https://github.com/DouglasBoubert/selectionfunctions}) produced by Boubert \& Everall (2020a, in prep) based on the {\sc dustmaps} source code \citep{dustmaps18}.

%%%%%%%%%%%%
\section{Conclusions}
\label{sec:conclusion}

We have developed a Bayesian model for empirically determining the selection function of multi-fibre spectrographs, where there exists a complete photometric survey in the same region of observable (colour-apparent magnitude) parameter space. 

The method improves on previous works by modelling the selection function with a Gaussian mixture model that is fit to the data through a Poisson likelihood function. This generates a selection function which accounts for Poisson noise in low-count data. This approach also allows us to define the uncertainties in our selection function by analysing the posterior distribution on the model parameters. 

We further incorporate a union calculation which allows the selection function to be calculated in regions of sky where fields partially or fully overlap. This can be applied to merged catalogues of independent surveys to produce a combined selection function. In an era where large amounts of spectroscopic data are becoming available from many independent observatories, each with their own observational limitations, combining surveys can hugely enhance our understanding of the Milky Way. We can also apply our method to subsamples of any catalogue if analysis is being done on a more specific or constrained stellar population.

Finally, we present a method of translating selection functions from observable colour-apparent magnitude coordinates into intrinsic coordinates of age, metallicity, mass and distance using the PARSEC isochrones \citep{PARSEC12}. This allows a deeper insight into the effects of the selection function on stellar parameters.

We have demonstrated the effectiveness of our method on a mock catalogue generated using the \texttt{Galaxia} \citep{Galaxia11} population synthesis code. We are successful in reproducing the applied selection function within Poisson noise uncertainty. Using KS tests in one dimension we show that our method produces good fits to the data only struggling where the model selection functions undergo large discontinuous transitions.

Our code is made publicly available as a {\sc Python} repository at \url{https://github.com/AndrewEverall/seestar.git} which will continue to be improved and updated with new developments. We will apply our method to several spectroscopic catalogues and make these selection functions publicly available.

\section*{Acknowledgements}

AE is grateful to the Galactic Dynamics group in Oxford for providing the support and funding necessary to develop this work. AE in particular thanks Prof James Binney, Dr John Magorrian, and Dr Ralph Sch\"{o}nrich for regular discussions, which led to many improvements. AE thanks Zephyr Penoyre and Stephen Thorp for useful discussions on optimisation techniques. We are extremely greatful to the referee who provided detailed comments which lead to significant improvements to the paper and the method.

%%%%%%%%%%%%%%%%%%%% REFERENCES %%%%%%%%%%%%%%%%%%%%%%

\bibliographystyle{mnras}
\bibliography{SFpaper_mnras}

%%%%%%%%%%%%%%%%%%%%%%%%%%%%%%%%%%%%%%%%%%%%%%%%%%

%%%%%%%%%%%%%%%%% APPENDICES %%%%%%%%%%%%%%%%%%%%%%%%%

\appendix

\section{Union expansion}
 \label{app:unionexp}
 
Considering the simplest case of two overlapping fields, A and B, Equation~\eqref{eq:fullunion} expands to
\begin{equation}
\label{eq:double}
\begin{split}
\Prob(\Se  \mid   \bth, \mathbf{v}) = \Prob(\Se_\mathrm{A} \mid  \bth, \mathbf{v}) +\Prob(\Se_\mathrm{B} \mid  \bth, \mathbf{v}) - \Prob(\Se_\mathrm{A}, \Se_\mathrm{B} \mid  \bth, \mathbf{v}).
\end{split}
\end{equation}
In most surveys we are considering, there will be substantially more than two fields. For $M$ fields, we expand the union using inclusion-exclusion principle.
\begin{equation}
\begin{split}
\Prob \left (\bigcup_{i=1}^M \Se_i \right) = &\sum_{i=1}^M \Prob(\Se_i) -\sum_{i=2}^M\sum_{j=1}^{i-1} \Prob(\Se_i, \Se_j)  \\
\label{eq:expansion}
&+\sum_{i=3}^M\sum_{j=2}^{i-1} \sum_{k=1}^{j-1}\Prob(\Se_i,\Se_j,\Se_k) - ... \\
 & ...  + (-1)^{M+1}  \Prob(\Se_1, \Se_2....\Se_M) \\
 = &\sum_{k=1}^M (-1)^{k+1} \left[   \sum_{1\leq i_1 < ... < i_k \leq M} \Prob(\Se_{i_1}, \Se_{i_2} ...\Se{i_k}) \right],
 \end{split}
\end{equation}
where we have dropped the conditionals such that {$\Prob(\Se_i) \equiv  \Prob(\Se_i \mid  \bth, \mathbf{v} )$} for ease of notation.

Assuming independence of observations from different fields (i.e. the event that a star is selected on field A is independent of whether it has been observed on field B) we can expand the joint probability as the product of the probability of each event.
\begin{equation}
\begin{split}
\Prob(\Se_1, \Se_2, ...\mid\bth, \mathbf{v}) = \prod_{i = 1, 2, ...}   \Prob(\Se_i \mid  \bth, \mathbf{v}) \nonumber
\end{split}
\end{equation}

Using Equation~\ref{eq:fieldselection} we expand the conditional selection probabilities in terms of the event that the positional coordinates lie on the field
\begin{equation}
\begin{split}
\Prob(\Se_1, \Se_2, ...\mid\bth, \mathbf{v}) = \prod_{i = 1, 2, ...}   \Prob(\Se_i \mid  \Theta_i, \mathbf{v})\,\Prob(\Theta_i \mid \bth) \nonumber
\end{split}
\end{equation}
where $\Prob(\Theta_i \mid \bth) =1$ if $\bth$ is on field i and $\Prob(\Theta_i   \mid\bth) = 0$ otherwise. Therefore any joint probability terms with fields which don't contain $\bth$ will vanish from Equation~\ref{eq:expansion}.

We can simplify Equation~\eqref{eq:expansion} for some specific circumstances:

\begin{enumerate}[label=(\arabic*)]
\item $\bth$ is not located on any fields. $\Prob(\Theta_i  \mid  \bth) = 0 \quad \forall \quad i$. All terms in the expansion are $0$
	$$ \Prob \left (\bigcup_{i=1}^M \Se_i\right) = 0 $$
\item $\bth$ is located on only one patch, $A$. 
$\Prob(\Theta_i  \mid  \bth) = 0 \quad \forall \quad i\neq A$, $\Prob(\Theta_A \mid  \bth) = 1$
	$$\Prob \left (\bigcup_{i=1}^M \Se_i\right) = \Prob(\Se_A \mid  \Theta_A, \mathbf{v}) $$
\item $\bth$ is on the intersection between two fields denoted by $A$ and $B$:
	\begin{align}
	\label{eq:double2}
	\Prob \left (\bigcup_{i=1}^M \Se_i\right) = &\Prob(\Se_A \mid  \Theta_A, \mathbf{v}) + \Prob(\Se_B \mid  \Theta_B, \mathbf{v})\\
						    &- \Prob(\Se_A \mid  \Theta_A, \mathbf{v}) \times \Prob(\Se_B \mid  \Theta_B, \mathbf{v})  \nonumber
	\end{align}
\end{enumerate}

\section{Poisson likelihood}
\label{app:poislike}

We start from the likelihood of observing a particular object with coordinates $v_i$ given that we only pick one point from the density function, $\lambda$
\begin{equation}
\Prob(v_i \mid n=1, \lambda) =\frac{ \lambda(v_i)}{\int \mathrm{d}v\,\lambda(v)}
\end{equation}
Expanding this to observations of a population of N objects
\begin{equation}
\Prob(v_1, v_2...v_N \mid n=N, \lambda) = \prod_{i=1}^N \frac{ \lambda(v_i)}{\int \mathrm{d}v\,\lambda(v)}
\end{equation}
The likelihood of the data is then given by
\begin{equation}
\begin{split}
\Prob(v_1, v_2...v_N, n=N & \mid \lambda) \\
=& \Prob(v_1, v_2...v_N \mid n=N, \lambda)\Prob(n=N \mid \lambda) \\
=&  \frac{\prod_{i=1}^N \lambda(v_i)}{(\int \mathrm{d}v\,\lambda(v))^N} \frac{(\int \mathrm{d}v\,\lambda(v))^N \exp{\left(-\int \mathrm{d}v\,\lambda(v)\right)}}{N!} \\
=&  \frac{\prod_{i=1}^N \lambda(v_i) \exp{\left(-\int \mathrm{d}v\,\lambda(v)\right)}}{N!}
\end{split}
\end{equation}
The denominator here is independent of the parameters of the density model so the likelihood of our dataset, $\{v\}$ is given by
\begin{equation}
\Prob(\left\{v\right\} \mid \lambda) \propto \prod_{i=1}^N \lambda(v_i) \exp{\left(-\int \mathrm{d}v\,\lambda(v)\right)}
\end{equation}
And taking the log likelihood
\begin{equation}
\ln\Prob(\left\{v\right\} \mid \lambda) \propto -\int \mathrm{d}v\,\lambda(v) + \sum_{i=1}^N \ln{\lambda(v_i)}
\end{equation}

\section{Model Structure.}
\label{app:modelapp}

In Table~\ref{tab:notation} we provide short descriptions of the notation followed in this paper to help with following the method.

\begin{table*}
\centering
\begin{tabular}{|| c | c || }
\hline
\hline
$\mathbf{x} = (\boldsymbol{\theta}, \mathbf{v})$ &  Full coordinates of the star.\\ 
$\left\{X_\mathrm{phot}\right\} (\left\{X_\mathrm{spec}\right\})$ &  Sample of photometric (spectroscopic) data.\\ 
$\boldsymbol{\theta}=(l,b)$ & Galactic coordinates. \\
$\mathbf{v} = (m, c)$ & Observable coordiantes: colour and apparent magnitude. \\
$\tau, \mathrm{[M/H]}, \mathcal{M}_\mathrm{ini}, s$ & Intrinsic coordinates: age, metallicity, initial mass and distance \\
$\widetilde{\mathcal{M}}_\mathrm{ini}$ & Initial mass scaled to the range $\widetilde{\mathcal{M}}_\mathrm{ini} \in [0,1]$.\\
\hline
$\Theta_i$ & Event that a star is on field $i$.\\ 
$\Se$ & Event that a star is selected in the spectroscopic catalogue.\\ 
$\Se_i$ & Event that a star is selected on field $i$ of the spectrosopic catalogue.\\
\hline
$f_\mathrm{phot} (f_\mathrm{spec}) $ & Distribution function (spectroscopic distribution). \\ 
$n_\mathrm{phot} (n_\mathrm{spec}) $ & Density of stars in the photometric (spectrosopic) sample.\\ 
$N_\mathrm{phot} (N_\mathrm{spec}) $ & Number of stars in the photometric (spectrosopic) sample.\\ 
\hline
$g_\mathrm{DF} (g_\mathrm{SF})$ & GMM for the distribution function (selection function)\\ 
$\mathbf{\epsilon} (\widetilde{\epsilon})$ & Parameters of the distribution function (selection function). \\ 
$\alpha_k$ & Dirichlet concentration parameters for the photometric density GMM.\\
$w_{i k}, \boldsymbol{\mu}_{i k}, \boldsymbol{\Sigma}_{i k}$ & Parameters of field $i$, component $k$ of the photometric density GMM.\\ 
$\pi_{i k}$ & Normalised component loadings of the photometric density GMM.\\ 
$N$ & Normalisation of the photometric density GMM.\\
$\widetilde{w_{i k}}, \widetilde{\boldsymbol{\mu}_{i k}}, \widetilde{\boldsymbol{\Sigma}_{i k}}$ & Parameters of field $i$, component $k$ of the selection function GMM.\\ 
$\widetilde{\kappa_{i k}}$ & logit transform of the component weights of the selection function GMM.\\ 
\hline
$\mathcal{L}^\mathrm{phot}(\mathcal{L}^\mathrm{spec})$ & Likelihood function for the photometric density (spectroscopic density) fit.\\
\hline
$M$ & Number of fields or patches.\\
$K$ & Number of GMM components in photometric density.\\
$\widetilde{K}$ & Number of GMM components in selection function.\\
\hline
\hline
\end{tabular}
\caption{Notation followed in this paper.}
\label{tab:notation}
\end{table*}

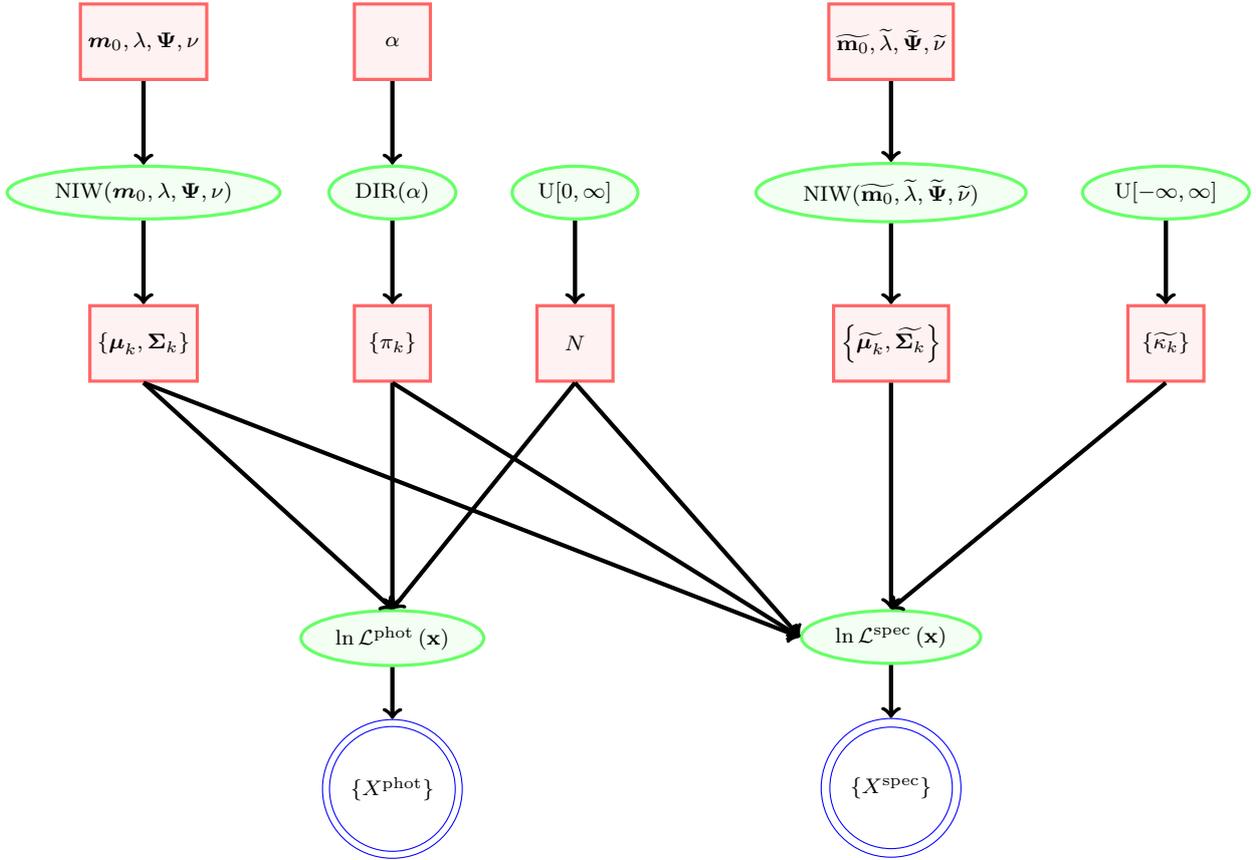
\begin {figure*}%[!hbtp]
\centering
%\begin{adjustbox}{width=\textwidth}
%\makebox[\textwidth]{
\resizebox{\width}{\height}{
\begin{tikzpicture}[nodes=draw,
roundnode/.style={ellipse, draw=green!60, fill=green!5, very thick, minimum size=7mm, minimum width=5mm},
squarednode/.style={rectangle, draw=red!60, fill=red!5, very thick, minimum size=10mm},
node distance=2cm and 4cm, 
]

%\draw [color=gray,fill=gray!5](-0.7,-0.8) rectangle (8.2,1);
%\node at (8.2,1) [below=2mm, left=1mm] {\textsc{Density profiles, Potential models + Poisson equation}};
%\draw [color=gray,fill=gray!5](-0.7,-3.1) rectangle (3.4,-1.5);
%\node at (3.4,-3.1) [above=2mm, left=1mm] {\textsc{Jeans Equations}};
%\draw [color=gray,fill=gray!5](-0.7,-5.5) rectangle (3.4,-3.7);
%\node at (3.4,-3.7) [below=2mm, left=1mm] {\textsc{Gaussian velocity}};
%\draw [color=gray,fill=gray!5](3.7,-5.5) rectangle (8.2,-3.7);
%\node at (8.2,-3.7) [below=2mm, left=1mm] {\textsc{Selection function}};
%\draw [color=gray,fill=gray!5](-1.5,-10.5) rectangle (11.2,-8.7);
%\node at (11.,-10.5) [above=2mm, left=1mm] {\textsc{Gaussian measurement uncertainties}};
%Nodes
	\node [squarednode, right=-3mm] (niw_hpphot) {$\boldsymbol{m}_0, \lambda, \boldsymbol{\Psi}, \nu$};
	\node [squarednode, right=1.9cm of niw_hpphot] (conc) {$\alpha$};
	\node [squarednode, right=5.2cm of conc] (niw_hpsf) {$\widetilde{\mathbf{m}_0}, \widetilde{\lambda}, \widetilde{\boldsymbol{\Psi}}, \widetilde{\nu}$};
	%\node [roundnode, below of=dmpot] (jeans) {$\overline{v}, \sigma_{ij}$};
	
	\node [roundnode, below of=conc] (dir_phot) {$\mathrm{DIR}(\alpha)$};
	\node [roundnode, below of=niw_hpphot] (niw_phot) {$\mathrm{NIW}(\boldsymbol{m}_0, \lambda, \boldsymbol{\Psi}, \nu)$};
	\node [roundnode, right=0.7cm of dir_phot] (Unorm_phot) {$\mathrm{U}[0, \infty]$};
	\node [roundnode, below of=niw_hpsf] (niw_sf) {$\mathrm{NIW}(\widetilde{\mathbf{m}_0}, \widetilde{\lambda}, \widetilde{\boldsymbol{\Psi}}, \widetilde{\nu})$};
	
	\node [squarednode, below of=dir_phot] (piphot) {$\{\pi_k\}$};
	\node [squarednode, below of=niw_phot] (musig_phot) {$\{\boldsymbol{\mu}_k, \boldsymbol{\Sigma}_k\}$};
	\node [squarednode, below of=Unorm_phot] (norm_phot) {$N$};
	\node [squarednode, below of=niw_sf] (musig_spec) {$\left\{\widetilde{\boldsymbol{\mu}_k}, \widetilde{\boldsymbol{\Sigma}_k}\right\}$};
	
	\node [roundnode, right=0.7cm of niw_sf] (conc_sf) {$\mathrm{U}[-\infty, \infty]$};
	\node [squarednode, below of=conc_sf] (pi_spec) {$\left\{\widetilde{\kappa_k}\right\}$};
	%, $\widetilde{w_k} = \sqrt{|2\pi\widetilde{\boldsymbol{\Sigma}_k}|}\,\mathrm{logit}^{-1}(\widetilde{\kappa_k})$
	
	\node [roundnode, below=3cm of piphot] (Lphot) {$\ln \mathcal{L}^\mathrm{phot} \left(\mathbf{x}\right)$};
	\node [roundnode, below=3cm of musig_spec] (Lspec) {$\ln \mathcal{L}^\mathrm{spec} \left(\mathbf{x}\right)$};
	
  	\node [draw=blue, shape=circle, inner sep=2.mm, below of=Lphot] () {$\{X^\mathrm{phot}\}$};
 	\node [draw=blue, shape=circle, inner sep=6.5mm, below of=Lphot] (phot_data) {};
  	\node [draw=blue, shape=circle, inner sep=2.mm, below of=Lspec] () {$\{X^\mathrm{spec}\}$};
 	\node [draw=blue, shape=circle, inner sep=6.5mm, below of=Lspec] (spec_data) {};
    
%Lines
\draw[->, ultra thick] (conc.south) -- (dir_phot.north);
\draw[->, ultra thick] (niw_hpphot.south) -- (niw_phot.north);
\draw[->, ultra thick] (niw_hpsf.south) -- (niw_sf.north);

\draw[->, ultra thick] (dir_phot.south) -- (piphot.north);
\draw[->, ultra thick] (niw_phot.south) -- (musig_phot.north);
\draw[->, ultra thick] (Unorm_phot.south) -- (norm_phot.north);
\draw[->, ultra thick] (niw_sf.south) -- (musig_spec.north);

\draw[->, ultra thick] (piphot.south) -- (Lphot.north);
\draw[->, ultra thick] (piphot.south) -- (Lspec.west);
\draw[->, ultra thick] (musig_phot.south) -- (Lphot.north);
\draw[->, ultra thick] (musig_phot.south) -- (Lspec.west);
\draw[->, ultra thick] (norm_phot.south) -- (Lphot.north);
\draw[->, ultra thick] (norm_phot.south) -- (Lspec.west);
\draw[->, ultra thick] (musig_spec.south) -- (Lspec.north);

%\draw[->, ultra thick] (musig_spec.south) -- (pi_spec.north);
\draw[->, ultra thick] (conc_sf.south) -- (pi_spec.north);
\draw[->, ultra thick] (pi_spec.south) -- (Lspec.north);

\draw[->, ultra thick] (Lphot.south) -- (phot_data.north);
\draw[->, ultra thick] (Lspec.south) -- (spec_data.north);
\end{tikzpicture}
}
\caption{Bayesian network (directed acyclic graph) of the model described in this paper as a function of colour and apparent magnitude for a single field. The network describes a method for determining the posterior of photometric density and selection function parameters where both are parameterised GMMs.}
%\end{adjustbox}
\label{fig:graph}
\end{figure*}

In Figure~\ref{fig:graph} we present a Bayesian network (or acyclic directed graph) of the method for calculating the selection functions for each field as a function of colour and apparent magnitude. Red boxes contain parameters and hyperparameters of the model. Green ellipses are conditional probabilities from which parameters and data are sampled. The blue double circles are the observable data.
In summary, the method consists of maximising the product off all probability distributions in green ellipses. The best fit hyperparameters then define the posterior parameters of the photometric density and selection function GMMs. In our method this is achieved as a two stage process, first fitting the photometric data to determine posterior GMM parameters for the photometric density. Subsequently the photometric GMM parameter posteriors are provided as priors to the spectroscopic likelihood function.

%%%%%%%%%%%%%%%%%%%%%%%%%%%%%%%%%%%%%%%%%%%%%%%%%%
% Don't change these lines
\bsp	% typesetting comment
\label{lastpage}
\end{document}